\documentclass[reprint,prb]{revtex4-2}

\usepackage{color}
\usepackage{amsmath,amsthm}
\usepackage{graphicx}
\usepackage{comment}
\usepackage{tikz}
\usetikzlibrary{decorations.pathreplacing}
\usetikzlibrary{calc}

\newcommand{\av}[1]{\left\langle #1 \right \rangle}
\newcommand{\abs}[1]{\left| #1 \right |}

\newcommand{\qv}{\mathbf{q}}
\newcommand{\kv}{\mathbf{k}}

\DeclareMathOperator{\sign}{sgn}

\begin{document}

\title{Dual Bethe-Salpeter equation for the multi-orbital lattice susceptibility\\ within dynamical mean-field theory}

\author{Erik G. C. P. van Loon}
\affiliation{NanoLund and Division of Mathematical Physics,
    Department of Physics,
    Lund University, Lund,
    Sweden}

\author{Hugo U. R. Strand}
\affiliation{School of Science and Technology, Örebro University, SE-701 82 Örebro, Sweden}
\affiliation{Institute for Molecules and Materials, Radboud University, 6525 AJ Nijmegen, the Netherlands}

\begin{abstract}
  Dynamical mean-field theory describes the impact of strong local correlation effects in many-electron systems.
  While the single-particle spectral function is directly obtained within the formalism,
  two-particle susceptibilities can also be obtained by solving the Bethe-Salpeter equation.
  The solution requires handling infinite matrices in Matsubara frequency space.
  This is commonly treated using a finite frequency cut-off, resulting in slow linear convergence.
A decomposition of the two-particle response in local and non-local contributions enables a reformulation of the Bethe-Salpeter equation inspired by the dual boson formalism. The re-formulation has a drastically improved cubic convergence with respect to the frequency cut-off, facilitating the calculation of susceptibilities in multi-orbital systems considerably. This improved convergence arises from the fact that local contributions can be measured in the impurity solver.
The dual Bethe-Salpeter equation uses the fully reducible vertex which is free from vertex divergences.
We benchmark the approach on several systems including the spin susceptibility of strontium ruthenate Sr$_2$RuO$_4$, a strongly correlated Hund's metal with three active orbitals. 
\end{abstract}

\maketitle
\section{Introduction}

Electronic correlations are present in many interesting quantum materials, including Mott insulators~\cite{Imada98}, ruthenates~\cite{Tamai19}, iron pnictides~\cite{Coldea18} and cuprates~\cite{Phillips22}. In these materials, the interaction between electronic quasiparticles is strong, leading to the breakdown of the independent quasi-particle picture. The theoretical description of this quantum many-body problem is hard and generally requires approximations.
The Dynamical mean-field theory (DMFT)~\cite{Metzner89, Georges96} approximation is one prominent example, which quantitatively explains many experimentally observed phenomena~\cite{Tamai19}, although there are also situations where it is insufficient~\cite{Rohringer18, Qin22}.

\begin{figure}
\includegraphics[scale=1]{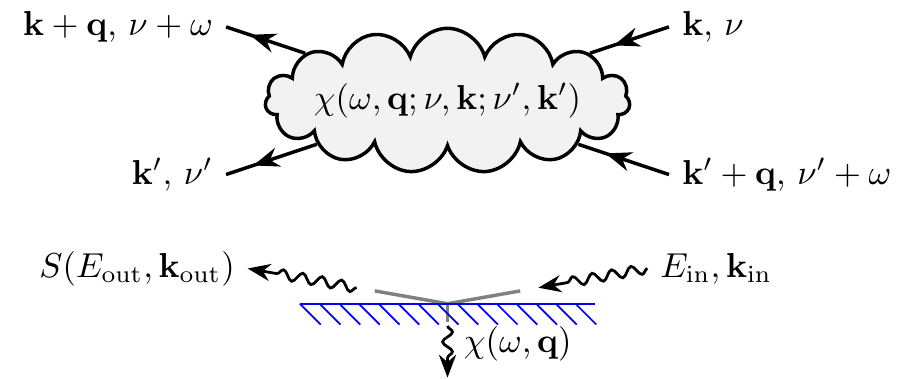}\ \\[-4mm]
\caption{Generalized susceptibility $\chi$ (top), schematic scattering experiment (bottom).}\ \\[-8mm]
\label{fig:expt}
\end{figure}

The properties of a system of correlated electrons can be described in terms of expectation values of the form  $\langle c^\dagger_\alpha c^{\phantom{\dagger}}_\beta \rangle$, $\langle c^{\dagger}_{\alpha}c^{\phantom{\dagger}}_{\beta}c^\dagger_{\gamma} c^{\phantom{\dagger}}_{\delta} \rangle$ and so on, where $c_\alpha$, $c_\alpha^\dagger$ stand for the annihilation and creation operator of an electron and $\alpha$ is a combined time $\tau$, space $\mathbf{R}$, and spin-orbital $a$ label $\alpha \equiv (a, \mathbf{R}, \tau)$. Here $\langle c^\dagger_\alpha c^{\phantom{\dagger}}_\beta \rangle$ is called a single-particle quantity, $\langle c^\dagger_{\alpha}c^{\phantom{\dagger}}_{\beta}c^\dagger_{\gamma} c^{\phantom{\dagger}}_{\delta} \rangle$ a two-particle quantity, and so on. With the increase in the number of operators and labels in the correlation function, these many-particle quantities become increasingly complicated. In practice, expectation values with more than two pairs of operators are often (but not always~\cite{Ribic17,Kappl22}) out of reach, but both the single-particle and two-particle quantities are experimentally accessible.
A simplification of the structure is possible for translational invariant systems in equilibrium where only relative time and position matters. Whence, the number of space-time (or momentum-frequency) labels in the correlation function can be reduced by one. Another name for the two-particle quantity is the generalized susceptibility $\chi$, see Fig.\ \ref{fig:expt}, since these correlation functions describe the linear response of the electronic system to an external field, according to the fluctuation-dissipation theorem.

Experimentally, the susceptibilities are accessible via techniques such as inelastic neutron scattering (INS) \cite{Enderle:2014}, electron energy loss spectroscopy (EELS)~\cite{Platzman73} or resonant inelastic x-ray scattering (RIXS) \cite{RevModPhys.83.705}. Roughly speaking, these experiments consist of shooting a probing particle with a known energy $E_\text{in}$ and momentum $\mathbf{k}_\text{in}$ at the correlated material of interest, and measuring the distribution $S$ over energy $E_\text{out}$ and momentum $\mathbf{k}_\text{out}$ of the outgoing probing particle. The energy and momentum difference between the ingoing and outgoing particle is transferred to the sample, i.e.\ the energy $\omega = E_\text{in} - E_\text{out}$ and momentum $\qv = \mathbf{k}_\text{in} - \mathbf{k}_\text{out}$ are absorbed by the correlated material, and large magnitudes of $S$ indicate that the material supports (collective) excitations at $(\omega,\qv)$. Thus, the scattering amplitude $S$ measured by these spectroscopic techniques is proportional to the susceptibility $\chi(\omega,\qv)$ of the material, up to further scattering matrix element and interaction effects, see Fig.\ \ref{fig:expt}.

Thus, the experimental spectrum $S$ only depends on the transferred energy and momentum $(\omega,\qv)$, while the generalized electronic susceptibility $\chi$ is a four-index object, which depends on three frequencies and momenta due to time-space translation symmetry. To get to the experimentally relevant susceptibility, two of the frequencies and momenta of the generalized susceptibility need to be traced out.
To understand the meaning of this trace, consider the momentum labels $\qv,\kv,\kv'$ of the generalized susceptibility. These labels correspond to an electronic transition $\kv\rightarrow \kv+\qv$ and another electronic transition $\kv'+\qv \rightarrow \kv'$, see Fig.\ \ref{fig:expt}. In the theoretical description, it is possible to consider/calculate the likelihood of such a transition as a function of both $\kv$ and $\kv'$, as well as $\qv$, but in the experiment, only the momentum transfer $\qv$ is observable and there is no way to determine the momenta of the electronic states involved in the transition, and the trace takes the form
\begin{equation*}
  S(E_\text{out}, \mathbf{k}_\text{out}) \propto \chi(\omega, \mathbf{q}) = \sum_{\mathbf{k} \mathbf{k}'} \sum_{\nu \nu'} \chi(\omega, \mathbf{q}; \nu, \mathbf{k}; \nu', \mathbf{k}')
  \, .
\end{equation*}
Thus, there is a tremendous loss of information when going from the full generalized susceptibility to a particular traced-out version that is actually experimentally observable.

In this work, we will show that this loss of information can actually be used to our advantage when we want to calculate the susceptibility in DMFT~\cite{Georges96}. The usual formulation of the Bethe-Salpeter equation in DMFT proceeds by first calculating the generalized susceptibility of the material, based on the generalized susceptibility of the auxiliary impurity model, and only taking the trace over momenta and fermionic frequencies at the end. The disadvantage of this is that this formulation of the Bethe-Salpeter equation converges only \emph{linearly} with respect to the number of internal frequencies used in the calculation, especially for local scattering processes. 
Some techniques have been developed to correct for the known high-frequency asymptotics~\cite{PhysRevB.83.085102,Boehnke:2011fk,Kaufmann17,Tagliavini18,Wentzell20,PhysRevB.101.035110}, but converging with respect to the frequency grid remains a challenge for reaching experimentally relevant and physically interesting temperatures.

An alternative formulation~\cite{Pruschke96}, which can be derived using the dual boson approach~\cite{Rubtsov12}, takes advantage of the distinction between local and non-local fluctuations in DMFT. By using the local single-frequency and two-frequency susceptibilities of the auxiliary impurity model (measured directly in the impurity solver), it is possible to efficiently take into account all local scattering processes, regardless of their internal frequency. The resulting dual Bethe-Salpeter equation (DBSE) then only has to account for non-local scattering processes~\cite{Hafermann14,vanLoon14,vanLoon15,vanLoon16,Krien19}, which converge \emph{cubically} with respect to the number of fermionic frequencies used in the calculation. This is a substantial improvement that enables calculations at lower temperature or in more complex systems, since the computational complexity of the DBSE (and BSE) scales with the number of orbitals $N_o$ and the momentum discretization $N_k$, as well as the number of fermionic frequencies $N_\nu$ according to $\mathcal{O}( N_{\nu}^3 N_{o}^6 N_k )$. 

This dual boson based approach has previously been applied to charge~\cite{vanLoon14PRL}, $S^z$~\cite{vanLoon16} and full spin~\cite{Krien17} susceptibility of single-orbital models, and a multi-orbital implementation of this scheme using a channel decomposition is available in the AbinitioD$\Gamma$A project~\cite{Galler19}. 
Here, we present an open-source implementation of the dual Bethe-Salpeter equation in the two particle response function toolbox (TPRF \cite{Strand:tprf}), based on the toolbox for interacting quantum systems (TRIQS~\cite{triqs}). We provide a detailed benchmark of the static spin susceptibility of Sr$_2$RuO$_4$. 

\section{Theory}

We will consider lattice models in thermodynamic equilibrium at finite temperature, using the imaginary time and Matsubara frequency formalism. We start with a brief overview of the susceptibility and how it is usually calculated in DMFT, including our notation for describing it. Some familiarity with these concepts is assumed, we refer the interested reader to the literature~\cite{Georges96,Pavarini2014dmft,BoehnkeThesis,Rohringer18,KrienThesis} for more details. A discussion of the relationship with previous works can be found in Sec.~\ref{sec:discussion}.

\subsection{Notation}

The single-particle Green's function is defined as $G_{a\bar{b}}(\tau,\mathbf{R}) = \av{c^{\phantom{\dagger}}_{\tau,a,\mathbf{R}} c^\dagger_{0,b,\mathbf{0}}} $, where $\tau\in [0,\beta)$ is the imaginary time, $a$ and $b$ are combined spin-orbital labels and $\mathbf{R}$ is a lattice vector. We use $\bar{\cdot}$ to denote indices that appear on creation operators, which becomes very relevant for many-particle Green's functions. The Fourier transform of $G_{a\bar{b}}(\tau,\mathbf{R})$ is $G_{a\bar{b}}(\nu,\kv)$, where $\nu$ is a fermionic Matsubara frequency and $\kv$ is the momentum in the first Brillouin Zone. 

Similarly, the generalized susceptibility $\chi_{\bar{a}b\bar{c}d}(\omega,\qv)$ is the Fourier transform of
\begin{align}
  \chi_{\bar{a}b\bar{c}d}(\tau,\mathbf{R})
  =& 
  \av{c^{\dagger}_{\tau,\mathbf{R},a} c^{\phantom{\dagger}}_{\tau,\mathbf{R},b} c^{\dagger}_{0,0,c} c^{\phantom{\dagger}}_{0,0,d} }
  \notag \\
  &-
  \av{c^{\dagger}_{\tau,\mathbf{R},a} c^{\phantom{\dagger}}_{\tau,\mathbf{R},b}}
  \av{c^{\dagger}_{0,0,c} c^{\phantom{\dagger}}_{0,0,d}} \notag \\
  =&
  \av{c^{\dagger}_{\tau,\mathbf{R},a} c^{\phantom{\dagger}}_{\tau,\mathbf{R},b} c^{\dagger}_{0,0,c} c^{\phantom{\dagger}}_{0,0,d} } \notag \\
  &-
  G_{b\bar{a}}(\beta, \mathbf{0})G_{d\bar{c}}(\beta, \mathbf{0})
  \, . \label{eq:chi:def}
\end{align}
Here, $\omega$ is a bosonic Matsubara frequency. For clarity, we use $\qv$ for the bosonic momentum and $\kv$ for fermionic momenta, both take values in the first Brillouin Zone.

Note that the order of barred operators differs between the single-particle Green's function and the susceptibility. The notation of the single-particle Green's function is conventional. For the susceptibility there are many notations found in the literature, we follow the notation of TPRF~\cite{Strand:tprf}, see also Appendix~\ref{app:conventions}.

The spin-orbital labels can be contracted to go from the generalized susceptibility to a particular physical susceptibility, $\chi^{AB}(\omega,\qv)= \sum_{abcd} A_{ab} B_{cd} \chi_{\bar{a}b\bar{c}d}(\omega,\qv)$, where $A$ and $B$ are matrices in spin-orbital space.
As an example, the $S^z$ component of the spin susceptibility is obtained as $\chi^{S^z S^z}(\omega,\qv)=\sum_{abcd} \hat{\sigma}^z_{ab} \hat{\sigma}^z_{cd} \chi_{\bar{a}b\bar{c}d}(\omega,\qv)$, where $\hat{\sigma}^z$ is the appropriate Pauli matrix. 

The static susceptibility $\chi^{AB}(\omega=0,\qv=0)$ can be interpreted in linear response theory. Define the operators $\hat{A}=\sum_{ab} A_{ab} c^\dagger_a c_b$ and $\hat{B}=\sum_{cd} B_{cd} c^\dagger_c c_d$, which we assume to be Hermitian. Then, the application of a spatially homogeneous field $\lambda \hat{B}$, in the sense that the Hamiltonian changes as $\hat{H}\rightarrow \hat{H}- \lambda \hat{B}$, leads to a change in $\langle \hat{A} \rangle$, where the zero field response is $d \langle \hat{A} \rangle/d\lambda \big|_{\lambda = 0}=\chi^{AB}(\omega=0,\qv=0)$. This definition can be generalized to spatially inhomogeneous fields (finite $\qv$) by considering enlarged unit cells~\cite{Strand19}, while finite $\omega$ corresponds to time dependent applied fields.

\subsection{Bethe-Salpeter equation in Dynamical Mean-Field Theory}

In DMFT, the susceptibility can be calculated using the Bethe-Salpeter equation \cite{Georges96}. Here, we introduce the equations without orbital, frequency and momentum labels to emphasize the structure of the equations, explicit formulas including labels are given later. To distinguish the physical susceptibility $\chi$ of the system from the auxiliary impurity quantities in DMFT, it is commonly called the \emph{lattice} susceptibility. In general, the Bethe-Salpeter equation for $\chi$ takes the form
\begin{equation}
  \chi = \chi^0 + \chi^0 \Gamma \chi, \label{eq:dmft-bse}
\end{equation}
where the bubble $\chi^0 = G \ast G$ denotes the independent propagation of a particle-hole pair according to the Green's function $G$ given by the DMFT approximation. $\Gamma$ is the irreducible vertex, which is assumed to be local in the DMFT approximation. Both $G$ and $\Gamma$ can be extracted from the solution of DMFT's auxiliary impurity problem. The Bethe-Salpeter Equation \eqref{eq:dmft-bse} is a self-consistent equation for $\chi$, which can be inverted to give $\chi$ as a function $\chi^0$ and $\Gamma$.

In DMFT, it is also possible to discuss the susceptibility of the auxiliary impurity problem, $X$, which has it's own Bethe-Salpeter equation 
\begin{equation}
  X = X^0 + X^0 \Gamma X. \label{eq:imp-bse}
\end{equation}
Here, the impurity bubble $X^0 = g\ast g$ is defined in analogy with the lattice bubble, and $g$ is the Green's function of the impurity model, see also Eq.\ \eqref{eq:X0}.
Importantly, within the DMFT approximation the lattice and impurity vertices $\Gamma$ are identical.
The frequency, momentum and orbital dependence of the products in Eqs.\ \eqref{eq:dmft-bse} and \eqref{eq:imp-bse} is here supressed for readability, c.f.\ Sec.\ \ref{sec:dualbse}.

\subsection{Locality}

Locality is a central concept in DMFT. The DMFT self-consistency condition guarantees that $g$ is the local part of $G$, which is often expressed as $g=G(\mathbf{R}=0)=\frac{1}{N_k}\sum_\mathbf{k} G(\mathbf{k})$. This implies that $X^0$ is the local part of $\chi^0$ as well, since the bubble is a direct product in real space $\chi^0(\tau, \mathbf{R}) = -G(\tau, \mathbf{R}) G(-\tau, -\mathbf{R})$, whence
\begin{equation*}
  \chi^0(\tau, \mathbf{0}) = -G(\tau, \mathbf{0}) G(-\tau, \mathbf{0}) = -g(\tau) g(-\tau) = X^0(\tau)
  \, .
\end{equation*}
However, $X$ is not the local part of $\chi$~\cite{vanLoon15}, since the Bethe-Salpeter equation \eqref{eq:dmft-bse} allows for processes where the electron-hole pair moves to a different site and back. Physically, $X$ corresponds to the linear response of the impurity model while keeping the dynamical mean-field fixed, whereas $\chi$ corresponds to linear response where the mean-field is self-consistently adjusted.

Although $X$ is not identical to (the local part of) $\chi$, it might still be a much better starting point for the calculation of $\chi$ than $\chi^0$, since $X$ already contains local vertex corrections in the form of Eq.~\eqref{eq:imp-bse}. We should stress that $X$ can be extracted directly from the impurity model, without needing to use the identity Eq.~\eqref{eq:imp-bse}. Thus, the central idea of the dual Bethe-Salpeter equation~\cite{Rubtsov12,Hafermann14} is to express $\chi$ as $X$ and some corrections.  
Finally, we note that this idea has also been used to decompose the polarization~\cite{Krien19}.

\subsection{Dual Bethe-Salpeter equation: idea}

As mentioned above, it makes sense to pull apart the lattice bubble $\chi^0$ into two parts, the impurity bubble $X^0$ and a remainder $\tilde{\chi}^0$, 
\begin{equation}
  \chi^0 = X^0 + \tilde{\chi}^0. 
\end{equation}
We are looking for an analogous separation of $\chi$,
\begin{equation}
\chi = X + \overline{\chi}
\end{equation}
in terms of the impurity susceptibility $X$ and a correction term $\overline{\chi}$, derived below. Insertion of the Bethe-Salpeter equations [Eqs.\ (\ref{eq:dmft-bse}) and (\ref{eq:imp-bse})] gives
\begin{align}
  \overline{\chi} &= \chi - X
  =
  \frac{\chi^0}{1 - \chi^0 \Gamma} - \frac{X^0}{1 - X^0 \Gamma}
  \\ &=
  \frac{X^0 + \tilde{\chi}^0 }{1 - (X^0 + \tilde{\chi}^0) \Gamma} - \frac{X^0}{1 - X^0 \Gamma}
  \\ &=
  \frac{1}{1 - X^0 \Gamma}
  \frac{ \tilde{\chi}^0 }{ 1 - \tilde{\chi}^0 \frac{\Gamma}{1 - \Gamma X^0 } }
  \frac{1}{1 - X^0 \Gamma}
  \, .
\end{align}
The result can be further simplified by writing it in terms of the impurity reducible vertex $F$, defined as
\begin{equation}
  X = X^0 + X^0 F X^0
  \, , \quad
  F = \Gamma + \Gamma X^0 F \label{eq:idea:X}
\end{equation}
which gives the relations
\begin{equation}
  \frac{\Gamma}{1 - \Gamma X^0 } = F
  \, , \quad
  \frac{1}{1 - X^0 \Gamma} = 1 + X^0 F \equiv L. \label{eq:idea:L}
\end{equation}
Here, we have introduced $L$ to compactify the notation further.

Hence, the DMFT+DBSE approach gives the lattice susceptibility $\chi$ in terms of the impurity susceptibility $X$ and the dual correction term
\begin{equation}
  \chi = X + L \frac{ \tilde{\chi}^0 }{ 1 - \tilde{\chi}^0 F } L \equiv X + L \tilde{\chi} L \label{eq:dual-bse-brief}
  \, ,
\end{equation}
see Eqs.\ \eqref{eq:dualbse} and \eqref{eq:susc:final} for the complete frequency, momentum and orbital structure.
Note that this formulation of the lattice susceptibility only uses the reducible vertex $F$ and is therefore free from the issues related to divergences that occur in $\Gamma$~\cite{Schafer13,Chalupa18,Reitner20,vanLoon20}.

Both formulations of the Bethe-Salpeter equation, Eqs.~\eqref{eq:dmft-bse} and \eqref{eq:dual-bse-brief}, are matrix equations, and the matrix multiplication involves a formally infinite sum over fermionic frequencies. In practice, this sum has to be cut off at some $N_\nu$. In Sec.~\ref{sec:convergence}, we will show that the Eq.~\eqref{eq:dual-bse-brief} converges more rapidly with respect to $N_\nu$, making it superior for practical implementations. Before studying the convergence, however, we will first introduce how the calculation is done in practice, with all relevant orbital and frequency labels and prefactors.

\subsection{Impurity correlators}
\label{sec:impurity_correlators}

To achieve the improved scaling with $N_\nu$, it is central that the three constituents of the right-hand side of the DBSE equation \eqref{eq:dual-bse-brief}, $X$, $L$ and $F$ are obtained from the impurity model directly, as described below, from independent Monte Carlo measurements. Note that, while it is possible to obtain the components with zero or one fermionic frequency ($X$ and $L$) using sums over frequencies of the generalized two-particle Green's function ($g^{(4)}$ defined below), doing so would introduce its own $N_\nu$-cutoff errors and would make the scaling with the size of the frequency box worse.

The impurity susceptibility $X_{\bar{a}b\bar{c}d}(\omega)$ is the Fourier transform to Matsubara frequency of Eq.~\eqref{eq:chi:def} for the impurity model, i.e., without a spatial label $\mathbf{R}$. 

Similarly, for the fermion-boson vertex $L$, we start with the impurity correlation function $g^{(3)}_{\bar{a}b\bar{c}d}(\omega,\nu)=\av{c^{\dagger}_{\nu,a} c^{\phantom{\dagger}}_{\nu+\omega,b} (c^\dagger_{c} c^{\phantom{\dagger}}_d)_\omega}$, where $\nu$ is a fermionic and $\omega$ a bosonic Matsubara frequency. $L$ is the amputated connected component of $g^{(3)}$, i.e.,
\begin{multline}
 -\frac{1}{T} g_{e\bar{a}}(\nu) g_{b\bar{f}}(\nu+\omega)  L_{\bar{e}f\bar{c}d}(\omega,\nu) \\
 = g^{(3)}_{\bar{a}b\bar{c}d}(\omega,\nu)-\frac{1}{T} g_{b\bar{a}}(\nu)g_{d\bar{c}}(\tau=0^+) \delta_{\omega,0}. \label{eq:L:amputate}
\end{multline}
Here, $g$ is the single-particle Green's function of the impurity model. To solve Eq.~\eqref{eq:L:amputate} for $L$, it is convenient to express it in terms of the impurity bubble,
\begin{equation}
X^0_{\bar{a}b\bar{c}d}(\omega,\nu,\nu')=-\frac{1}{T} \delta_{\nu\nu'} g_{d\bar{a}}(\nu)g_{b\bar{c}}(\nu+\omega),
\label{eq:X0}
\end{equation}
so that Eq.~\eqref{eq:L:amputate} becomes 
\begin{multline}
 X^0_{\bar{a}b\bar{f}e}(\omega,\nu,\nu)  L_{\bar{e}f\bar{c}d}(\omega,\nu) \\
 = g^{(3)}_{\bar{a}b\bar{c}d}(\omega,\nu)-\frac{1}{T} g_{b\bar{a}}(\nu)g_{d\bar{c}}(\tau=0^+) \delta_{\omega,0} \label{eq:L:amputate:2}.
\end{multline}
This is a tensorial equation in orbital space while being diagonal in both the fermionic and the bosonic Matsubara frequency. $L$ is obtained by multiplication with the inverse of $X^0$, TPRF's implementation of these operations is used and this is the reason for the factor $-1/T$ in our definition of $L$. 

The fermion-boson vertex has some useful properties, which have been discussed in detail for single-orbital DMFT~\cite{vanLoon18fermionboson}. First, the impurity susceptibility $X$ is equal to $g^{(3)}$ traced over $\nu$. Secondly, $L$ describes the linear response of the self-energy to an external field, 
\begin{equation}
  -\frac{1}{T}\sum_{cd} B_{\bar{d}c} L_{\bar{a} b \bar{c} d}(\omega=0, \nu) = B_{\bar{a}b} +  \frac{\partial \Sigma_{b\bar{a}}(\nu)}{\partial \lambda }. \label{eq:L:linearresponse}
\end{equation}
Thirdly, for a non-interacting impurity model, $L_{\bar{a} b \bar{c} d}(\omega,\nu) = - T\delta_{\bar{a}d}\delta_{\bar{c}b}$, is constant, since $-Tg^{(3)}_{\bar{a}b\bar{c} d}(\omega,\nu)=g_{d\bar{a}}(\nu) g_{b\bar{c}}(\nu+\omega)-g_{b\bar{a}}(\nu)g_{d\bar{c}}(\tau=0^+) \delta_{\omega,0}$. This expression for $L$ is consistent with the linear response of Eq.~\eqref{eq:L:linearresponse}. It can also be shown more generally that $L$ asymptotically goes to a constant in the limit of large $\nu$, keeping $\omega$ fixed. 

Finally, $L$ has a symmetry with respect to complex conjugation. In general,
\begin{align}
 L_{\bar{a}b\bar{c}d}(\omega,\nu-\omega/2)^\ast = L_{\bar{b}a\bar{d}c}(\omega,-\nu-\omega/2) \label{eq:L:symmetry}.
\end{align}
For real Hamiltonians, i.e., a Hamiltonian where all matrix elements are real in the chosen orbital representation, which includes all examples studied here, there is the additional symmetry relation
\begin{align}
 L_{\bar{a}b\bar{c}d}(\omega,\nu-\omega/2) \overset{\text{real }\hat{H}}= L_{\bar{b}a\bar{d}c}(-\omega,\nu+\omega/2) \label{eq:L:symmetry:real}.
\end{align}
This formula is obtained by reversing the arrow of all fermionic lines (time-reversal), and re-arranging the diagram to bring it back into the customary form. These formulas are proven in Appendix~\ref{app:L:symmetry}.

From our definition in Eq.~\eqref{eq:L:amputate}, $L$ has units of energy since $T$ has units of energy, $g(\nu)$ has units of inverse energy and $g(\tau)$ is dimensionless. For the linear response, it is convenient to define the operators $\hat{A}$ and $\hat{B}$ in a dimensionless way (this includes the important examples of $\hat{N}$ and $\hat{S}^i$), so $\lambda$ has units of energy and the right-hand side of Eq.~\eqref{eq:L:linearresponse} is dimensionless. 

Another ingredient for the dual Bethe-Salpeter equation is the reducible impurity vertex, $F_{\bar{a}b\bar{c}d}$, which is the connected, amputated two-particle correlation function of the impurity model. This object is used in many DMFT-based theories~\cite{Rohringer18}, and a variety of notations is used in the literature. We use the same convention as TPRF for the particle-hole channel, i.e.,
\begin{multline}
  g^{(4)}_{\bar{a}b\bar{c}d}(\omega,\nu,\nu') =
  \\
  g_{e\bar{a}}(\nu) g_{b\bar{f}}(\nu \! + \! \omega)
  F_{\bar{e}f\bar{g}h}(\omega,\nu,\nu')
  g_{g\bar{c}}(\nu' \! + \! \omega) g_{d\bar{h}}(\nu')
  \\
  + \beta \delta_{0, \omega} g_{b\bar{a}}(\nu) g_{d\bar{c}}(\nu')
  - \beta \delta_{\nu, \nu'} g_{d\bar{a}}(\nu) g_{b\bar{c}}(\nu + \omega)
  \, .
  \label{eq:Fimp}
\end{multline}
In practice $g$ and $g^{(4)}$ are obtained from the impurity solver and $F$ is calculated using this relation by subtracting the disconnected components from $g^{(4)}$ and then amputating the four attached single particle Green's function legs by applying four $g^{-1}$ terms.

\subsection{Dual Bethe-Salpeter equation in multi-orbital systems}
\label{sec:dualbse}

The lattice Green's function $G$ in DMFT is defined by
\begin{equation}
 \delta_{a\bar{c}} = \left[ i\nu \delta_{a\bar{b}} - t_{a\bar{b}}(\mathbf{k}) - \Sigma_{a\bar{b}}(\nu) \right] G_{b\bar{c}}(\nu,\mathbf{k}),
\end{equation}
where $\Sigma$ is the self-energy of the auxiliary impurity model. We define the non-local single-particle Green's function $\tilde{G}$ as
\begin{equation}
  \tilde{G}_{a\bar{b}}(\nu, \mathbf{R}) = \begin{cases}
                                              G_{a\bar{b}}(\nu, \mathbf{R}) &\text{ if }\mathbf{R}\neq 0 \\
                                              0 &\text{ if }\mathbf{R}=0
  \end{cases},
  \label{eq:g_tilde_k}
\end{equation}
where $G_{a\bar{b}}(\nu, \mathbf{R})$ is the Fourier transform of $G_{a\bar{b}}(\nu, \mathbf{k})$ to real space. Corresponding to the non-local Green's function $\tilde{G}$, there is a ``bubble'' $\tilde{\chi}^0$,
\begin{equation}
  \tilde{\chi}^0_{\bar{a}b\bar{c}d}(\omega, \nu, \nu',\mathbf{R}) = - \frac{1}{T} \delta_{\nu \nu'} \tilde{G}_{d \bar{a}}(\nu,\mathbf{R}) \tilde{G}_{b \bar{c}}(\nu + \omega,-\mathbf{R}). \label{eq:dualbubble}
\end{equation}
This object is Fourier transformed back to momentum space, giving $\tilde{\chi}^0_{\bar{a}b\bar{c}d}(\omega, \nu, \nu',\mathbf{q})$. Evaluating the bubble in real space and Fourier transforming is more efficient than doing the calculation directly in momentum space~\cite{Hafermannphd,Kaltakphd}. We should note that $\tilde{\chi}^0$ is the nonlocal part of $\chi^0$ and $\tilde{\chi}^0(\omega,\nu,\nu',\mathbf{R}=\mathbf{0})=0$. Since both Green's functions in Eq.~\eqref{eq:dualbubble} have the same $\abs{\mathbf{R}}$, there are no cross-terms with one local and one non-local part of the Green's function.

From $\tilde{\chi}^0$ and $F$, the dual or non-local Bethe-Salpeter equation
\begin{align}
  &\tilde{\chi}_{\bar{a}b\bar{c}d}(\omega, \nu, \nu',\mathbf{q}) =
  \tilde{\chi}^0_{\bar{a}b\bar{c}d}(\omega, \nu, \nu',\mathbf{q}) +
  \label{eq:dualbse} \\
  &\,\,\,\,
  \tilde{\chi}^0_{\bar{a}b\bar{f}e}(\omega, \nu, \nu_1,\mathbf{q})
  F_{\bar{e}f\bar{g}h}(\omega,\nu_1,\nu_2)
  \tilde{\chi}_{\bar{h}g\bar{c} d}(\omega, \nu_2, \nu',\mathbf{q})
  , \notag
\end{align}
can now be solved for $\tilde{\chi}$ by inversion, using the reducible vertex $F$ obtained from Eq.\ \eqref{eq:Fimp}.

\begin{figure}
 \begin{tikzpicture}
 
 \node[anchor=east] at (1,0) {$L_{\bar{a}b\bar{c}d}(\omega,\nu)=$} ;
 
  \draw[black,thick,fill=red] (2,0) -- ++(-0.5,-0.5) -- ++ (0,1) -- cycle;  
  
  \node[below] at (1.5,-0.5) {$b$};
  \node[below,yshift=-10pt] at (1.5,-0.5) {$\nu+\omega$};
  \draw (1.5,-0.5) -- node {$>$} ++(-0.4,0) ;
  
  \node[above] at (1.5,+0.5) {$a$};
  \node[above,yshift=+10pt] at (1.5,+0.5) {$\nu$};
  \draw (1.5,+0.5) -- node {$<$} ++(-0.4,0) ;
  
  \draw (+2.5,-0.3) -- node[sloped] {$>$} (2,0) ;
  \draw (+2.5,+0.3) -- node[sloped] {$<$} (2,0) ;
  
  \node[right] at (+2.5,+0.3) {$d$} ;
  \node[right] at (+2.5,-0.3) {$c$} ;
  
  \draw [very thick, decorate,decoration={brace,mirror,amplitude=5pt},xshift=15pt] (2.5,-0.3) -- node[right,xshift=5pt] {$\omega$} (2.5,0.3) ;

 \end{tikzpicture} \\
 \begin{tikzpicture}
  
  \draw[black,thick,fill=red] (0,0) -- ++(0.5,-0.5) -- ++ (0,1) -- cycle;

  \draw[black,thick,fill=red] (2,0) -- ++(-0.5,-0.5) -- ++ (0,1) -- cycle;
  
  \draw[black,fill=blue!20] (0.5,-0.5) -- node {$>$} (1.5,-0.5) -- (1.5,0.5) -- node {$<$} (0.5,0.5) -- cycle ;
  
  \node at (1,0) {$\tilde{\chi}$} ;
  \node[below] at (0.5,-0.5) {$f$};
  \node[below] at (1.5,-0.5) {$g$};
  \node[below,yshift=-10pt] at (0.5,-0.5) {$\nu\phantom{'}+\omega$};
  \node[below,yshift=-10pt] at (1.5,-0.5) {$\nu'+\omega$};
  
  \node[above] at (0.5,0.5) {$e$};
  \node[above] at (1.5,+0.5) {$h$};
  \node[above,yshift=+10pt] at (0.5,+0.5) {$\nu$};
  \node[above,yshift=+10pt] at (1.5,+0.5) {$\nu'$};
  
  \draw (-0.5,-0.3) -- node[sloped] {$>$} (0,0) ;
  \draw (-0.5,+0.3) -- node[sloped] {$<$} (0,0) ;

  \draw (+2.5,-0.3) -- node[sloped] {$>$} (2,0) ;
  \draw (+2.5,+0.3) -- node[sloped] {$<$} (2,0) ;
  
  \node[left] at (-0.5,+0.3) {$a$} ;
  \node[left] at (-0.5,-0.3) {$b$} ;
  \node[right] at (+2.5,+0.3) {$d$} ;
  \node[right] at (+2.5,-0.3) {$c$} ;
  
  \draw [very thick, decorate,decoration={brace,amplitude=5pt},xshift=-15pt] (-0.5,-0.3) -- node[left,xshift=-5pt] {$-\omega,-\mathbf{q}$} (-0.5,0.3) ;

  \draw [very thick, decorate,decoration={brace,mirror,amplitude=5pt},xshift=15pt] (2.5,-0.3) -- node[right,xshift=5pt] {$\omega,\mathbf{q}$} (2.5,0.3) ;

 \end{tikzpicture}
 \caption{Diagrammatic representation of elements of the dual Bethe-Salpeter equation. At the top, the orbital and frequency definition of $L$ is illustrated. In our definition, $L(\omega,\nu)$ implies that the incoming electron \emph{emits} energy $\omega$ to the bosonic degree of freedom. At the bottom, the non-local contribution to the susceptibility is drawn. In this process, energy $\omega$ and momentum $q$ comes in from the left, is transferred to an electron-hole pair that propagates non-locally ($\tilde{\chi}$) and is then emitted. In the triangle on the left, since the electron \emph{absorbs} energy, the corresponding argument of the first $L$ in Eq.~\eqref{eq:susc:final} is $-\omega$.}
 \label{fig:diagrams}
\end{figure}
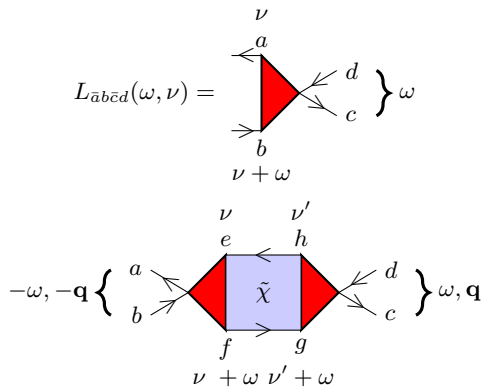

Finally, to get the physical susceptibility, we have to combine the impurity susceptibility with the non-local correction, adding fermion-boson vertices at the end, as sketched in Fig.~\ref{fig:diagrams},
\begin{align}
  &\chi_{\bar{a}b\bar{c}d}(\omega,\mathbf{q}) = X_{\bar{a}b\bar{c}d}(\omega)
  \notag \\ 
  &\,\,\,\,
  +
  L_{\bar{f}e\bar{a}b}(-\omega,\nu+\omega) \tilde{\chi}_{\bar{e}f\bar{g}h}(\omega,\nu,\nu',\mathbf{q}) L_{\bar{h}g\bar{c}d}(\omega,\nu')
  . \label{eq:susc:final}
\end{align}

Using the symmetry Eq.~\eqref{eq:L:symmetry} of $L$, this can be rewritten as
\begin{align}
  &\chi_{\bar{a}b\bar{c}d}(\omega,\mathbf{q}) = X_{\bar{a}b\bar{c}d}(\omega)
  \notag \\ 
  &\,\,\,\,
  +
  L^*_{\bar{e}f\bar{b}a}(-\omega,-\nu) \tilde{\chi}_{\bar{e}f\bar{g}h}(\omega,\nu,\nu',\mathbf{q}) L_{\bar{h}g\bar{c}d}(\omega,\nu')
  . \label{eq:susc:final2}
\end{align}
For real Hamiltonians, using the symmetry Eq.~\eqref{eq:L:symmetry:real} of $L$, this can be rewritten as
\begin{align}
  &\chi_{\bar{a}b\bar{c}d}(\omega,\mathbf{q}) \overset{\text{real $\hat{H}$}}= X_{\bar{a}b\bar{c}d}(\omega)
  \notag \\ 
  &\,\,\,\,
  +
  L_{\bar{e}f\bar{b}a}(\omega,\nu) \tilde{\chi}_{\bar{e}f\bar{g}h}(\omega,\nu,\nu',\mathbf{q}) L_{\bar{h}g\bar{c}d}(\omega,\nu')
  . \label{eq:susc:final3}
\end{align}
Equations~\eqref{eq:susc:final2} and \eqref{eq:susc:final3} perform better in terms of frequency windows when $\omega$ is large, since $\nu+\omega$ is replaced by $\pm \nu$.
For real Hamiltonians, Eq.~\eqref{eq:susc:final3} also has the advantage that it only uses a single bosonic frequency $\omega$.
On the other hand, Equation~\eqref{eq:susc:final} is obviously invariant under an orbital basis transformation (counting barred and non-barred occurences of orbital labels on both sides of the equation), while this is not obvious to see in Equations~\eqref{eq:susc:final2} and \eqref{eq:susc:final3}. 

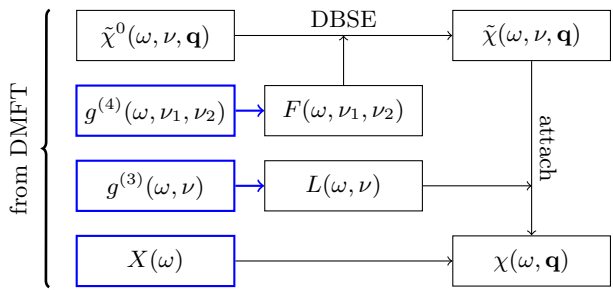
\begin{figure}
\begin{tikzpicture}[
block/.style={
draw,
rectangle, 
minimum width={width("$g^{(4)}(\omega,\nu_1,\nu_2)$")+6pt},
minimum height={19pt},
font=\small}]
  
 \node[block] at (2,2) (x0dual) {$\tilde{\chi}^{0}(\omega,\nu,\mathbf{q})$} ;
 \node[block,draw=blue,thick] at (2,1) (g4) {$g^{(4)}(\omega,\nu_1,\nu_2)$} ;
 \node[block,draw=blue,thick] at (2,0) (g3) {$g^{(3)}(\omega,\nu)$} ;
 \node[block,draw=blue,thick] at (2,-1) (X) {$X(\omega)$} ;

 \node[block] at (4.5,0) (L) {$L(\omega,\nu)$} ;
 \node[block] at (4.5,1) (F) {$F(\omega,\nu_1,\nu_2)$} ;

 \node[block] at (7,2) (xdual) {$\tilde{\chi}(\omega,\nu,\mathbf{q})$} ;

\node[block] at (7,-1) (chi) {$\chi(\omega,\mathbf{q})$} ;

 \draw[->,blue,thick] (g3) -- (L) ;
 \draw[->,blue,thick] (g4) -- (F) ;
 
 \draw[->] (X) -- (chi) ;
 
 \draw[->] (x0dual) -- node[above] {DBSE} (xdual) ;
 \draw[->] (F) -- ($(x0dual)!0.5!(xdual)$);
 
 \draw[->] (xdual) -- node[above,sloped] {attach} (chi) ;
 \draw[->] (L) -- ($(L)+(2.5,0)$) ;
 
 \draw[decorate,decoration={brace},line width=1pt] ([xshift=-10pt]X.south west) -- node[above,sloped,yshift=5pt] {from DMFT} ([xshift=-10pt]x0dual.north west) ;
 
\end{tikzpicture}
\caption{Flowchart summarizing the calculation. The blue boxes are local two-particle correlation functions extracted from the impurity solver. For clarity, orbital labels are suppressed here. }
\label{fig:flowchart}
\end{figure}

The entire procedure is summarized in Fig.~\ref{fig:flowchart}.
The computationally non-trivial step in this procedure is the inversion of the DBSE in Eq.~\eqref{eq:dualbse}. It corresponds to a geometric series with kernel $\tilde{\chi}^0 F$, which is divergent if one of the eigenvalues of the kernel is equal to $+1$, signaling a phase transition. Thus, analysis of this kernel can give valuable insight into the leading electronic fluctuations~\cite{vanLoon20}.

\subsection{Comparison of dual and usual Bethe-Salpeter equation}

The usual Bethe-Salpeter equation in DMFT is based on the Green's function $G$ instead of $\tilde{G}$, and the vertex $\Gamma$ instead of $F$. The susceptibility is then obtained directly as the output of the Bethe-Salpeter equation, i.e., without needing Eq.~\eqref{eq:susc:final}. Diagrammatically, it corresponds to the repeated scattering of electron-hole pairs propagating with the DMFT Green's function and interacting with the irreducible vertex $\Gamma$.

Comparing the two formulations of the Bethe-Salpeter equation, the dual boson based approach can be interpreted as a resummation of the original Bethe-Salpeter equation where terms in the geometric series are grouped together if subsequent vertices $\Gamma$ lie at the same lattice site. This occurs in the following way: If all vertices $\Gamma$ involve only a single lattice site, the diagram is a part of $X$. Otherwise, all vertices before the first change of site are grouped into the $L$ on the left and all vertices after the last change of site are grouped into the $L$ on the right. In between, all vertices $\Gamma$ that occur between two changes of site are grouped into a single $F$. To avoid double-counting of diagrams, the dual bubble $\tilde{\chi}^0$ is used to enforce a change of sites between every vertex. 

This construction provides for a one-to-one correspondence of the original Bethe-Salpeter series and the new, dual Bethe-Salpeter with all elements expressed in terms of $\Gamma$ and bubbles. In the derivation, $X$ and $L$ are expressed using local Bethe-Salpeter equations, \eqref{eq:idea:X}-\eqref{eq:idea:L}, but in practice they are obtained directly from the impurity solver, see Fig.~\ref{fig:flowchart}. This eliminates cutoff errors associated with the local Bethe-Salpeter equation.

\subsection{Implementation details}

After converging an ordinary DMFT self-consistency loop using TRIQS and the CTHYB solver~\cite{cthyb}, we use  w2dynamics~\cite{w2dynamics} to measure the two-particle correlation functions with one, two and three frequencies.
Due to different conventions the output of the w2dynamics solver is then transformed to the TPRF two-particle convention, see Appendix~\ref{app:conventions} for details.
After post-processing the two-particle correlation functions to get $X$, $L$ and $F$, the dual Bethe-Salpeter equation is evaluated and $\chi(\omega,\mathbf{q})$ is extracted as in Sec.~\ref{sec:dualbse}. This evaluation is the main contribution of the present manuscript.

The DMFT+DBSE calculation is compute bound by the measurements of the impurity correlation functions (using QMC).
The dual Bethe-Salpeter equation solver is memory bound, since we use fast Fourier transforms to evaluate the lattice bubble susceptibility $\chi^0$ as direct products in real space and imaginary time ($\mathbf{R}$, $\tau$).
However, the DBSE equation is diagonal in $\omega$ and $\mathbf{q}$ and can be evaluated one $(\omega,\mathbf{q})$ at a time.
All computationally non-trivial parts of the DBSE calculation are performed using the two-particle response function toolbox (TPRF) \cite{Strand:tprf} with routines implemented in \texttt{C++}, using hybrid OpenMP+MPI parallelization. 

\section{Convergence with respect to the number of fermionic frequencies}
\label{sec:convergence}
 
An important benefit of the dual version of the susceptibility calculation is the improved convergence with respect to the number of fermionic frequencies $N_\nu$. Whereas the traditional Bethe-Salpeter equation scales as $N_\nu^{-1}$, the new implementation generically scales as $N_\nu^{-3}$, as discussed below (see also \cite{Krien19}). We discuss the general case first and then give a few examples that are analytically tractable. 

\subsection{Cubic scaling of DBSE}

Since we are interested in the formal asymptotic scaling for large fermionic frequency, we eventually have $\omega \ll \nu$ and we can set $\omega=0$ for the analysis. In practice, for finite $\omega$, the asymptotic scaling sets in at larger $N_\nu$, and one typically needs $N_\nu \gtrsim \beta \omega / (2 \pi)$. In the analysis below, we drop the orbital labels, which are not relevant for the asymptotic analysis.

At large fermionic frequency, the single-particle Green's function behaves as $G(i\nu,\mathbf{k}) \sim \frac{1}{i\nu} + O(\nu^{-2})$. In other words, the leading term is $\mathbf{k}$-independent and hence completely local, and non-local parts of the Green's function decay at least as $(i\nu)^{-2}$. Thus, $\tilde{G}(i\nu,\mathbf{k})$ from Eq.~(\ref{eq:g_tilde_k}) decays as $(i\nu)^{-2}$ by construction~\cite{Rubtsov08}, since it is the Green's function with the local part removed. The dual bubble, Eq.~\eqref{eq:dualbubble} thus scales as $\tilde{\chi}^{0}(\omega=0,\nu,\nu,\mathbf{q}) \sim (i\nu)^{-4}$. It will turn out that this intermediate result is similar to the final result, so it is useful to have a look at the cutoff error here. With a fermionic frequency box of size $N_\nu$, the contribution from frequencies outside the box scales as $\sum_{\nu > N_\nu} \tilde{\chi}^{0}(\nu) \sim \sum_{\nu >N_\nu} (i\nu)^{-4} \sim \int_{N_\nu}^\infty dx/x^4 \sim N_\nu^{-3}$, i.e.\ the anticipated cubic scaling. Of course, the DBSE expression for the susceptibility is more complicated than just the dual bubble, so we need to analyze the effect of vertex corrections.

In the final expression for the susceptibility, Eq.~\eqref{eq:susc:final}, $X$ is independent of $N_\nu$, since it is measured directly in the impurity solver. The vertex $L(\omega,\nu)$, which (through $g^{(3)}$) is also measured in the impurity solver, is asymptotically constant in $\nu$~\cite{vanLoon18fermionboson}, so we can focus our attention on $\tilde{\chi}$, Eq.~\eqref{eq:dualbse}.
The vertex $F$ is asymptotically constant as a function of the fermionic Matsubara frequency~\cite{Rohringer12,Wentzell20}. The Bethe-Salpeter equation can be expanded as a geometric series, where we only write fermionic frequency labels,
\begin{multline}
\tilde{\chi}(\nu,\nu') = \tilde{\chi}^0(\nu)\delta_{\nu\nu'}+\tilde{\chi}^0(\nu) F \tilde{\chi}^0(\nu') \\
+ \sum_{\nu_1} \tilde{\chi}^0(\nu) F \tilde{\chi}^0(\nu_1) F \tilde{\chi}^0(\nu')+\ldots \, . 
\label{eq:scaling:bse} 
\end{multline}
For the dependence of the susceptibility on the fermionic frequency box size $N_\nu$, we are interested in the error $\sum_{\nu\nu'} \tilde{\chi}(\nu,\nu') - \sum_{\nu\nu' \leq N_\nu} \tilde{\chi}^{N_\nu}(\nu,\nu')$, where $\tilde{\chi}^{N_\nu}$ denotes that the fermionic frequency sums on the right-hand side of Eq.~\eqref{eq:scaling:bse} are also cut off at $N_\nu$. 
The first term decays as $\delta_{\nu\nu'} (i\nu)^{-4}$, which leads to $N_\nu^{-3}$ as before. For the second term, we need to distinguish between three cases~\cite{Tagliavini18,Wentzell20} for the cut-off error: (1) $\nu<N_\nu$, $\nu'>N_\nu$ (2) $\nu>N_\nu$, $\nu'<N_\nu$ and (3) $\nu>N_\nu$, $\nu'>N_\nu$. For (1), we know that $\tilde{\chi}^0 \sim (i\nu)^{-4}$ so the asymptotic contribution is of order $N_\nu^{-3}$. (2) is similar by symmetry. For (3), we have an error $\int_{N_\nu}^\infty d\nu (i\nu)^{-4} \int_{N_\nu}^\infty d\nu' (i\nu')^{-4} \sim N_\nu^{-6}$. Going to the third term in \eqref{eq:scaling:bse} or even higher order terms, more cases need to be distinguished, since it is now also possible for intermediate frequencies to be large. The leading contributions are always of order $N_\nu^{-3}$ and come from the cases where precisely one fermionic frequency is larger than $N_\nu$. Cases with more large fermionic frequencies have corresponding higher powers of $N_\nu^{-3}$.  

For the traditional BSE, the analysis proceeds in a very similar way, but the difference is that $\chi^0$ is used instead of $\tilde{\chi}^0$, and it scales as $\nu^{-2}$ only, leading to a cutoff error of $N_\nu^{-1}$. The same scaling holds for the local BSE. In fact, one can see that the trivial $\frac{1}{i\nu}$ asymptote of the Green's function is responsible for the slow scaling of the BSE. However, it contributes in a non-trivial way: the BSE is a series of ladder diagrams, and all orders in the series contribute to the $N_\nu$-asymptote via processes where a single rung of the ladder has $\nu>N_\nu$. 

\subsection{Example: The non-interacting dimer}

Consider the Hubbard dimer in the limit $U=0$, i.e., two atoms with a hopping amplitude $t$. In that case, the Hamiltonian has two eigenvalues $\epsilon(k)=\pm t$ and all calculations can be done straightforwardly. Define $\sign(k)=\sign(\epsilon_k)$, so $\epsilon_k=t \sign(k)$. 
\begin{align}
 G(i\nu,k) &= \frac{1}{i\nu - t\sign(k)}, \\
 \frac{1}{N_k}\sum_k G(i\nu,k) &= \frac{1}{2} \left( \frac{1}{i\nu - t} + \frac{1}{i\nu + t} \right), \\
 &= -\frac{i\nu}{\nu^2+t^2} \\
 \tilde{G}(i\nu,k) &= \sign(k) \frac{t}{\nu^2+t^2}.
\end{align}
The non-local bubble susceptibility at $q=0$ and $i\omega=0$ is 
\begin{align}
 \tilde{\chi}^0(\omega=0,\nu,\nu') &= -\beta \delta_{\nu\nu'} \frac{1}{N_k}\sum_k \tilde{G}(\nu,k)\tilde{G}(\nu,k) \\
 &= -\beta \delta_{\nu\nu'} \frac{t^2}{(\nu^2+t^2)^2} \\
 &= -\beta \delta_{\nu\nu'} \frac{t^2}{\nu^4} \frac{1}{(1+t^2/\nu^2)^2}.
\end{align}
From the final expression, it is clear that this frequency-diagonal matrix element decays as $(i\nu)^{-4}$. Given a fermionic frequency box of size $N_\nu$, the estimated total contribution from frequencies outside of the box scales as $\int_{N_\nu}^\infty dx/x^4 \propto N_\nu^{-3}$. Thus, we find cubic convergence.

For the dimer, it is easy to verify that the other value of $q$ has the same magnitude and only differs in the sign. Thus, it also has the same scaling. 

For completeness, the conventional evaluation of the susceptibility in the dimer, at $\omega=0$ and $q=0$, involves
\begin{align}
 \chi^0(\nu,\nu') &= -\beta \delta_{\nu\nu'} \frac{1}{N_k} \sum_k G(\nu,k)G(\nu,k) \\
 &= -\beta \delta_{\nu\nu'} \frac{1}{2} \left[\frac{1}{(i\nu-t)^2}+\frac{1}{(i\nu+t)^2}\right] \\
 &= -\beta \delta_{\nu\nu'} \frac{-\nu^2+t^2}{(\nu^2+t^2)^2}.
\end{align}
This matrix element decays as $(i\nu)^{-2}$, so the total cutoff error scales as $N_\nu^{-1}$, i.e., linearly.

The difference between $\chi^0$ and $\tilde{\chi}^0$ is given by the impurity susceptibility, which satisfies
\begin{align}
X &= -\beta\sum_\nu \left(\frac{1}{N_k}\sum_k G(i\nu,k)\right)^2 \\
&= -\beta\sum_\nu \frac{-\nu^2}{(\nu^2+t^2)^2}. 
\end{align}
However, this fermionic frequency sum is not performed explicitly in the fast evaluation of the susceptibility. Instead, $X$ is measured directly in the impurity solver.

\subsection{Tight-binding lattice model}

\begin{figure}
\ \\[-5mm]
\includegraphics{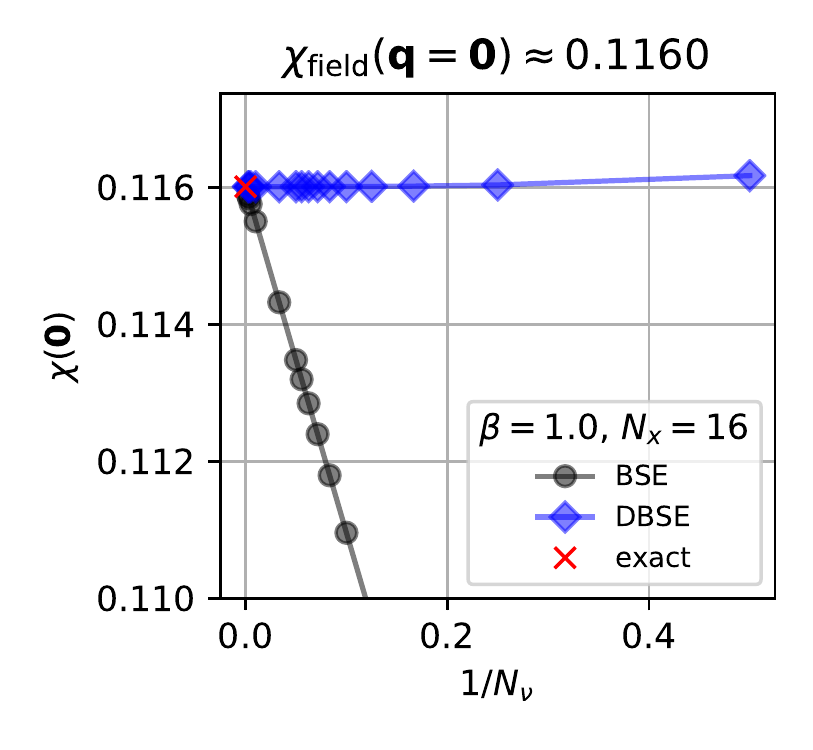}\ \\[-5mm]
\includegraphics{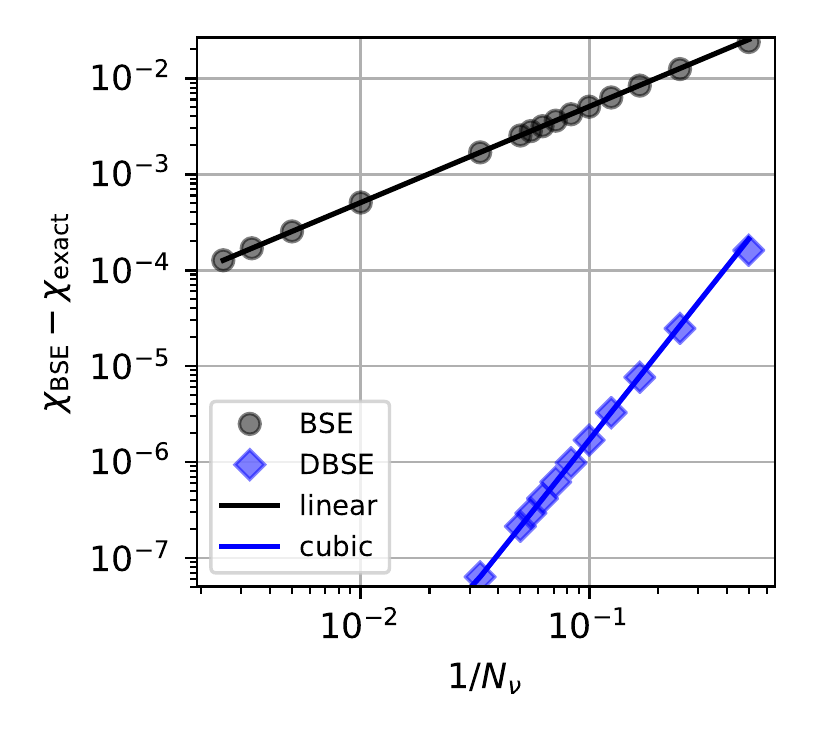}\ \\[-8mm]
\caption{Magnetic susceptibility at $\mathbf{q}=0$, $\omega=0$ in a two-orbital square lattice tight-binding model ($t=1$, $U=0$, $\mu\approx -1.84$) as a function of the number of frequencies $N_\nu$. The dual Bethe-Salpeter equation (DBSE) converges as $1/N_\nu^3$, whereas the usual Bethe-Salpeter equation (BSE) converges only as $1/N_\nu$, as is visible in the log-log plot on the right.}
\label{fig:U0}
\end{figure}

The scaling in a tight-binding lattice model (the non-interacting Hubbard model) is the same as in the corresponding non-interacting dimer, with the difference that the momentum sums cannot be done analytically. A numerical illustration of the scaling is given in Figure~\ref{fig:U0} for a square lattice with two orbitals per site. Note that the non-interacting model with orbital-diagonal hopping is actually just a sum over independent single-orbital models, so the orbital physics is trivial, but it is a useful test of the general implementation.
As a reference, Figure~\ref{fig:U0} uses the exact $\chi(\mathbf{q}=0,\omega=0)=dm/dh$ obtained using finite differences from the magnetization $m(h)$, given by the Fermi-Dirac distribution for non-interacting fermions. To achieve consistency, the same momentum space discretization should be used in the BSE and the linear response. 

\subsection{Atomic limit}
\label{sec:atomic}

In the atomic limit, $t_\kv=0$, we have $\chi(\omega,q)=X(\omega)$ and the new formulation gives the exact result regardless of the number of fermionic frequencies. Indeed, the dual bubble vanishes in the atomic limit, $\tilde{\chi}^0=0$. On the other hand, the conventional formulation of the Bethe-Salpeter equation does depend on the finite box size in the usual linear way, since the local Bethe-Salpeter equation of the Hubbard atom is non-trivial~\cite{Thunstrom18}.

\section{Benchmarks}

To benchmark the dual BSE formulation, we apply it to two strongly correlated models, namely the single-band Hubbard model on the square lattice and the three-band effective low-energy model for strontium ruthenate (Sr$_2$RuO$_4$).

\subsection{The single-band Hubbard model}

\begin{figure}
\includegraphics[scale=1]{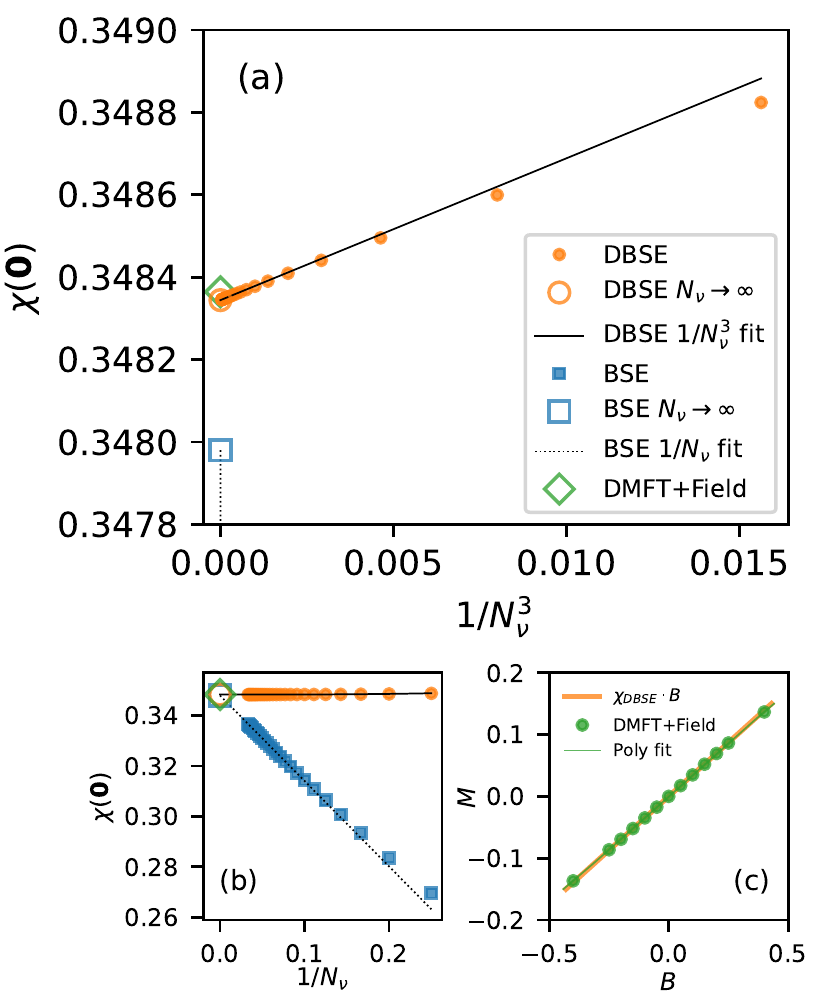}
\caption{Static magnetic susceptibility $\chi(\mathbf{0}, 0)$ ($\mathbf{q}=0$, $\omega=0$) in the square lattice Hubbard model as a function of the number of frequencies $N_\nu$. The dual BSE result (DBSE) (orange) is compared with the standard BSE (blue) and the linear response result from self-consistent DMFT in an applied field extrapolated to zero field (DMFT+Field) (green). Panel (a): Infinite frequency extrapolation (open markers) of calculations for several values of $N_\nu$ (filled markers) with $1/N_\nu^3$ frequency scaling on the x-axis showing the DBSE asymptotic $1/N_\nu^3$ scaling. Panel (b): Same as panel (a) but with $1/N_\nu$ scaling on the x-axis showing the BSE asymptotic $1/N_\nu$ scaling. Panel (c): Magnetization $M$ curve in applied magnetic field $B$ from self-consistent DMFT, used for zero-field extrapolation (green).}
\label{fig:fermi_hubbard}
\end{figure}

The single band Hubbard model on the square lattice is a canonical system for the study of strong correlations. Here we benchmark the static spin susceptibility, computed using dual BSE, against the standard BSE and the extrapolated zero field response from self consistent DMFT calculations in an applied magnetic field.
The calculations are performed in the strongly correlated regime with nearest neighbor hopping $t=1$ and local Hubbard interaction $U=10$ in the paramagnetic state at inverse temperature $\beta = 1$.

Figure~\ref{fig:fermi_hubbard}a shows the convergence of the DBSE static susceptibility (orange markers) using fermionic frequency windows in the range $4 \le N_\nu \le 30$, giving five digits of accuracy and a quantitative agreement with the applied field result (green diamond).
In fact, the discrepancy between the extrapolated DBSE result (orange open circle) and the applied field (green diamond) can be attributed to MC errors and the extrapolation to zero field in the linear response calculation (as shown in Fig.~\ref{fig:fermi_hubbard}c). This should be contrasted with the BSE results in Fig.~\ref{fig:fermi_hubbard}b where the raw data (blue squares) only give one digit of accuracy, requiring extrapolation to $N_\nu \rightarrow \infty$ to achieve any quantitative agreement (blue open square).

The expected cubic scaling with the size of the frequency window $N_\nu$ is clearly visible in the DBSE result (Fig.~\ref{fig:fermi_hubbard}a) allowing us to reach a higher accuracy than the conventional BSE with linear convergence.

\subsection{Three-band Hubbard model for strontium ruthenate}

\begin{figure}
 \includegraphics{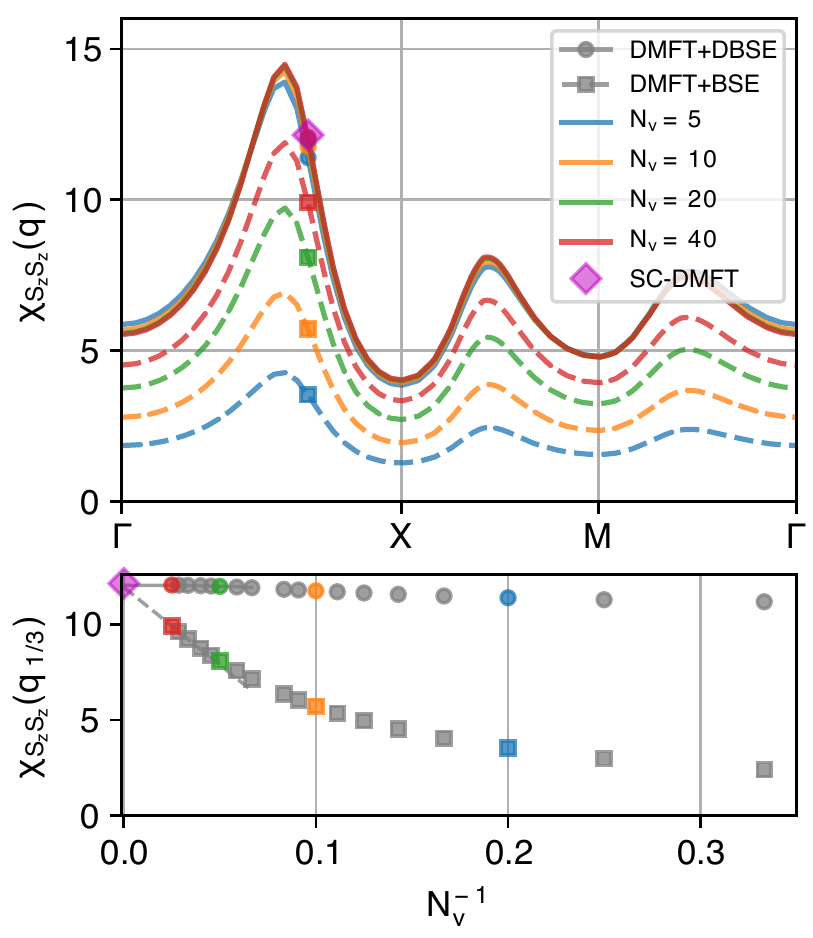}
 \caption{Upper panel: Static spin susceptibility $\chi_{S_z S_z}(\mathbf{q})$ of strontium ruthenate Sr$_2$RuO$_4$ at $T = 464$~K along the pseudo-tetragonal high-symmetry path ($\Gamma$-$X$-$M$-$\Gamma$). Results for $N_\nu = 5$, 10, 20, and 40 fermionic frequencies are shown for the DMFT+BSE method (dashed lines) and our dual BSE method (DMFT+DBSE) (solid lines). Lower panel: Convergence close to the incommensurate peak at $\mathbf{q} = (1/3, 1/3, 0)$ as a function of $1/N_\nu$ for $3 \le N_\nu \le 40$. Extrapolations to the infinite frequency limit $N_\nu \rightarrow \infty$ are also shown (lines), using the $N_\nu^{-3}$ scaling (solid line) for the DMFT+DBSE results (circles) and the $N_\nu^{-1}$ scaling (dashed line) for the DMFT+BSE results (squares). As an independent reference also the zero-field extrapolated self-consistent DMFT result from Ref.~\onlinecite{Strand19} is shown (magenta diamond marker).
 }
\label{fig:srvo:chi}
\end{figure}

While computing the lattice susceptibility within DMFT for the single band Hubbard model to higher accuracy is a feat, the main motivation for implementing the DBSE is to push the boundaries of applicability of the method to material realistic models in order to connect to experimental spectroscopies. This generally requires DMFT treatment of multi-orbital models. For example, the correlated electrons in transition metal compounds can often be described by effective low-energy Hubbard models with two to five orbitals per unit cell.

One such strongly correlated material is strontium ruthenate (Sr$_2$RuO$_4$) which has a substantial momentum-dependence in its spin susceptibility due to the presence of incommensurate spin fluctuations, as seen in INS experiments~\cite{Sidis99, Steffens19} and reproduced by DMFT+BSE calculations~\cite{0295-5075-122-5-57001, Acharya:2019aa, Strand19}.
The combination of orbital and momentum structure is responsible for the observed phenomena, with both strong local fluctuations and sharp features in momentum space. This makes Sr$_2$RuO$_4$ suitable as a benchmark for our dual Bethe-Salpeter equation implementation. Here, we focus on the frequency convergence of the implementation, the physics of the spin susceptibility in this material is discussed at length in Ref.~\onlinecite{Strand19}.




Following Ref.\ \onlinecite{Strand19}, we construct a low-energy model of Sr$_2$RuO$_4$ from the three bands crossing the Fermi level with Ru-4d t$_{2g}$ symmetry. The corresponding Wannier model was fit using Wannier90 \cite{Mostofi:2008aa} on a $10^3$ grid to the density functional theory band structure calculated with GPAW \cite{Mortensen2005, Enkovaara_2010}, using the PBE exchange correlation functional \cite{PhysRevLett.77.3865}, with a $20^3$ k-point grid, $600$ eV plane wave cutoff, and $25$ meV Methfessel-Paxton smearing \cite{PhysRevB.40.3616}, using the experimental crystal structure of Sr$_2$RuO$_4$ at $100\,$K \cite{PhysRevB.52.R9843}.
The interaction in the low energy three band t$_{2g}$ model was modeled using the rotationally invariant Kanamori interaction, with Hubbard $U=2.3$ eV and Hund's $J = 0.4$ eV and the self-consistent dynamical mean field calculations were performed at $T = 464$ K using TRIQS~\cite{triqs} and TRIQS/cthyb~\cite{cthyb} on a $16^3$ k-point grid. The two-particle quantities were measured using W2Dynamics \cite{WALLERBERGER2019388} worm sampling \cite{PhysRevB.92.155102} using $4 \cdot 10^9$ samples and a total compute budget of 110'000 core hours on the EuroHPC Joint Undertaking system LUMI.
The static spin-susceptibility $\chi_{S_z S_z}(\mathbf{q}, \omega=0)$ was finally computed on the level of DMFT+BSE and DMFT+DBSE using the Two-Particle Response Function toolbox TPRF \cite{Strand:tprf} using a $48^3$ k-point grid.

Figure~\ref{fig:srvo:chi} shows the resulting susceptibility along a high-symmetry path through the Brillouin Zone. Results are shown for both the traditional implementation of the Bethe-Salpeter equation (BSE) and our dual implementation (DBSE). In both cases, relatively small frequency boxes are used ($N_\nu=5,10,20,40$). For the conventional BSE, the slow $N_\nu^{-1}$ convergence is observed, and extrapolation is needed to obtain qualitatively correct results. On the other hand, the DBSE yields qualitative results already for the smallest box size of $N_\nu=5$, which is a truly remarkable improvement. Enlarging the frequency box, there is a small enhancement of the momentum-dependence, but no overall qualitative change. This is in stark contrast with the conventional BSE, where enlarging the fermionic frequency box increases the susceptibility across the entire Brillouin Zone. The origin of this difference is clear: local spin fluctuations are the major contribution to $\chi(\mathbf{q})$ and the dual Bethe-Salpeter equation includes all of these in one go via the impurity susceptibility, independent of the size of the fermionic frequency box.

For the computational efficiency of the approach, the cost of obtaining the two-particle correlation functions of the impurity model is also important. For this particular case, the worm sampling~\cite{PhysRevB.92.155102} used here (as implemented in w2dynamics) is orders of magnitude more efficient than the partition function sampling used in Ref.~\cite{Strand19}. Together, the improved convergence of the dual Bethe-Salpeter equation and the efficiency of the worm sampling make the calculation of DMFT lattice susceptibilities in multi-orbital models a routine task instead of a gargantuan undertaking.  

\subsection{Three-band Hubbard model for strontium vanadate}

\begin{figure}
 \includegraphics{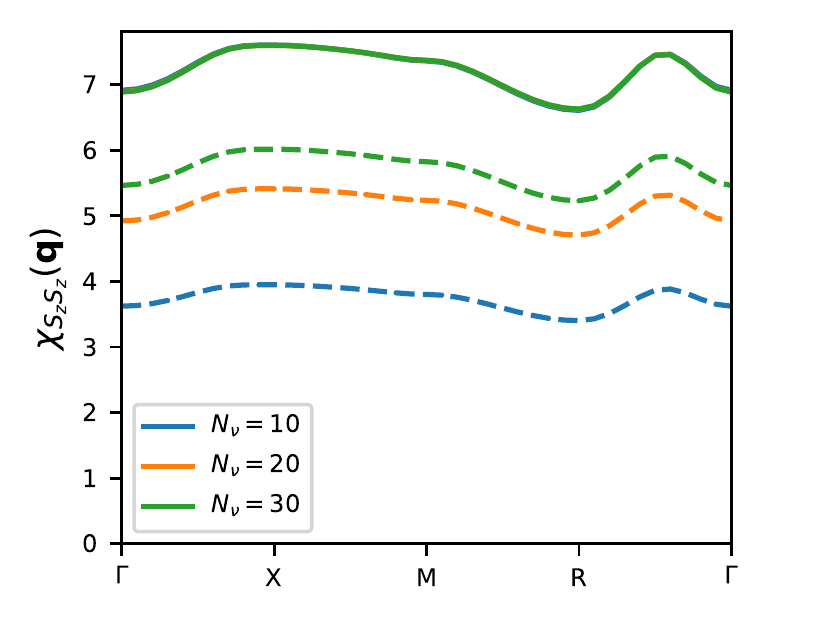} 
 \includegraphics{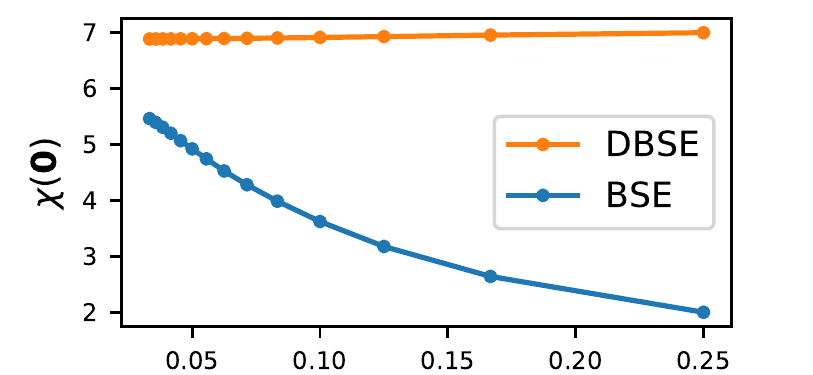} 
 \caption{Magnetic susceptibility of SrVO$_3$.  
 Upper panel: Susceptibility along a path between high-symmetry points, the solid lines are the DBSE and the dashed lines are the BSE. These calculations are performed with the Hamiltonian of Galler et al.~\cite{Galler19}, where the susceptibility was also calculated using the BSE (cf. their Fig. B 19), although their code is also capable of using the DBSE.  
 Lower panel: fermionic frequency convergence of the BSE and DBSE.  
 }
 \label{fig:srvo3}
\end{figure}

Strontium vanadate (SrVO$_3$) is another famous correlated material~\cite{Nekrasov06,Sakuma13}, with potential application as a transparent conductor~\cite{zhang2016correlated}. Galler et al.~\cite{Galler17,Galler19} have previously investigated the lattice susceptibility of SrVO$_3$ and provided a tight-binding Hamiltonian and reference data for the DMFT susceptibility~\cite{Galler19}, which makes it a useful test case for our current implementation. We have performed a self-consistent DMFT calculation based on their Hamiltonian ($U=5$, $J=0.75$ eV, $\beta=10$ eV, $20\times 20\times 20$ k-points), calculated the necessary impurity correlation functions using w2dynamics~\cite{w2dynamics} and subsequently calculated the lattice susceptibility using our implementation of the BSE and DBSE. Figure~\ref{fig:srvo3} shows the obtained magnetic susceptibility at $\omega=0$. For $N_\nu=30$, our results obtained using the BSE are consistent with Fig. B.19 of Ref.~\cite{Galler19}, which uses the same frequency box. We also performed susceptibility calculations at several $N_\nu \leq 30$ using our code and that of Ref.~\cite{Galler19}, and found relative deviations of approximately 2\% for the magnetic susceptibility at $\mathbf{q}=0$. This is quite small given that we are comparing calculations based on independent Monte Carlo solver runs, and that the details of the implementation differ.

As before, Figure~\ref{fig:srvo3} shows that the frequency convergence of the DBSE is so efficient that there is no visual difference between $N_\nu=10$, 20, 30, for a parameter set where the usual BSE has difficulty to reach convergence. Our converged DBSE results are qualitatively consistent with the results in Fig. 6 of Ref.~\cite{Galler19}, where a much larger frequency box $N_\nu=200$ was used. Thus, also for SrVO$_3$, we find that the DBSE greatly improves the convergence.

\section{Discussion}
\label{sec:discussion}

Here, we provide an overview of the previous literature and of similarities and differences between our implementation and other recent works. 

Early works showed that the locality of the self-energy in infinite dimensions also leads to simplifications on the two-particle level~\cite{Khurana90,Schweitzer91,PhysRevLett.69.168}. This led to a decomposition of the susceptibility into local and nonlocal parts~\cite{Pruschke96} similar to the formulas implemented here. 

A new view on momentum-resolved, dynamic susceptibilities was given by the dual boson method~\cite{Rubtsov12} for correlated systems. It was shown~\cite{Hafermann14} that the dual boson expression for the susceptibility is fully equivalent to the usual DMFT+BSE susceptibility under appropriate assumptions:  the equivalence holds when the interaction is fully local, no non-local self-energy diagrams are included in the dual theory and the same impurity model is used in both approaches. This is an example of the more general phenomenon that DMFT is included in the wider set of dual approximations~\cite{Rubtsov08,Brener08,Brener20}.

The dual formulation of DMFT, which makes a clear distinction between local and non-local processes, has certain advantages. Two major motivations for dual theories~\cite{Rubtsov08} are the faster decay with frequency and the good behavior in the strongly correlated limit. Regarding the fast decay, the anti-commutation relation $\{ c^{\phantom{\dagger}}_a,c^\dagger_b \}=\delta_{ab}$ determines the $(i\nu)^{-1}$-coefficient of the Green's function, which is therefore one for the local, diagonal part of the Green's function and zero otherwise. For the calculation of the susceptibility, the faster decay of the Green's function directly leads to a faster decay of the non-local bubble $\tilde{\chi}^0$ and forms the basis of the proof of cubic convergence in Sec.~\ref{sec:convergence}. For the second point, we have seen in Sec.~\ref{sec:atomic} that the non-local contribution to the susceptibility vanishes in the atomic limit, which suggests that this contribution will generically be small in strongly correlated systems. This is indeed seen in the large momentum-independent background of the magnetic susceptibility in Sr$_2$RuO$_4$ (Fig.~\ref{fig:srvo:chi} and Ref.~\cite{Strand19}) and SrVO$_3$ (e.g., Figure B.19 in Ref.~\cite{Galler19} and our Fig.~\ref{fig:srvo3}). 

A similar decomposition of the susceptibility based on local and non-local contributions was implemented in the AbinitioD$\Gamma$A~\cite{Galler17,Galler18,Galler19} code by Galler et al.~\cite{Galler19}. 
The AbinitioD$\Gamma$A code is set-up to use SU(2) symmetry and a decomposition into a charge and magnetic channel, but should be able to output full rank-4 susceptibilities with minor modifications.
This AbinitioD$\Gamma$A code is capable of reading measured versions of the impurity susceptibility and fermion-boson vertex, and should have improved frequency convergence in that case. However, in their example calculations, the one-frequency and two-frequency correlation functions are obtained by tracing out frequencies in $g^{(4)}(\omega,\nu,\nu')$, which does not lead to the improved convergence. For example, their Figure 6 and Figure B.19 are calculated with with 200 and 30 fermionic frequencies and show clear difference, a sign that the calculation is not converged at 30. 

Krien~\cite{Krien19} has presented a further improvement of the dual boson scheme based on the polarization instead of the susceptibility. The essential point is that in the present formulation, both $\tilde{\chi}^0$ and the vertex corrections have cutoff errors at the $1/N_\nu^3$ level. For the vertex corrections, these originate in processes where there is a single high-energy fermion propagator in the ladder. These attach to the vertex $F$, which is asymptotically constant and given by impurity processes which are reducible with respect to the bare interaction $\hat{U}$. By moving from the susceptibility to the polarization, this type of $\hat{U}$-reducible diagrams is to be excluded, so the associated $\hat{U}$-irreducible vertex $F^i$ decays asymptotically, instead of being constant. Thus, high-order diagrams in the Bethe-Salpeter series converge with additional powers of $N_\nu$. As a whole, Krien's formulation still depends on the cutoff as $1/N_\nu^3$, but only via $L\tilde{\chi}^0 L$. By using a larger frequency cut-off for $L$ than for $F$, which is beneficial since $L$ is typically much cheaper to calculate, one can thus reduce the impact of frequency cutoffs further. Krien's work deals with the single-orbital Hubbard model using charge and spin channels, but a generalization of his approach to generic rank-4 interaction tensors seems possible. Since the structure of this approach is similar, the basic tools provided by current implementation of the DBSE in TPRF could be used to implement Krien's formula: one needs to replace the input by their irreducible counterparts, i.e., $F$ by $F^i$, $L$ by $L^i$ and $X$ by $X^i=\pi(\omega)$, then one calculates the polarization via a DBSE structurally similar to the current one, and finally one transforms the obtained polarization to a susceptibility.

We should note that the improved frequency scaling we obtain is due to the dual Bethe-Salpeter reformulation in combination with a simple frequency cut-off. The improved scaling originates from applying the frequency cut-off on a different set of two-particle objects, with faster decaying high frequency asymptotics, than the terms in the standard DMFT Bethe-Salpeter equation. Regarding the general treatment of the Bethe-Salpeter equation, going beyond the drastic hard frequency cut-off in Matsubara space, there are several interesting approaches that incorporate the asymptotic frequency dependence in different ways~\cite{PhysRevB.83.085102,Boehnke:2011fk,Kaufmann17,Tagliavini18,Wentzell20,PhysRevB.101.035110} or that use a compact representation of the generic two frequency behavior \cite{PhysRevB.97.205111, PhysRevResearch.3.033168}.
Generally, the dual Bethe-Salpeter equation would benefit from these refined approaches, further improving the convergence.

\section{Conclusion}

We have presented an efficient implementation of the DMFT susceptibility for general multi-orbital systems based on the dual boson formalism. This algorithm reduces the scaling with the inverse number of frequencies from linear to cubic, leading to a dramatic speed-up of the calculation. A comparison with linear response shows that convergence with respect to the frequency box is much easier to achieve, leading to several digits of accuracy in the resulting static susceptibility in our example, even before extrapolation with respect to the size of the frequency box.

The implementation of the dual Bethe-Salpeter equation is based on the TRIQS library \cite{triqs} (in particular TPRF \cite{Strand:tprf}) and can in principle be used together with any impurity solver that provides the required two-particle correlation functions. An interface for the w2dynamics solver is provided with the implementation. We have found the worm sampling of two-particle correlation functions in w2dynamics to be very efficient for the susceptibilities studied here. Compared to the usual Bethe-Salpeter equation, one needs to compute two additional impurity correlation functions with zero and one fermionic frequencies, respectively. These are generally similar in computational cost to the vertex with two fermionic frequencies that is needed in both versions of the Bethe-Salpeter equation. However, Monte Carlo sampling issues have been observed for situations where the observables in the susceptibility do not commute with the Hamiltonian~\cite{vanloon2023larmor}. The difficulties associated with the solution of the impurity model remain the main bottleneck in the calculation of susceptibilities, and the more efficient convergence of the Bethe-Salpeter equation reduces the requirements on that bottleneck.

This efficient implementation of the Bethe-Salpeter equation can be useful for the study of magnetic~\cite{hausoel2017local,Strand19} and other susceptibilities~\cite{Geffroy19,Pickem21} in systems with multiple correlated orbitals in the unit cell. In particular, the better scaling with respect to the number of fermionic Matsubara frequencies makes it possible to reach lower temperatures at a given computational cost. 

\acknowledgments

EvL acknowledges support from Gyllenstiernska Krapperupsstiftelsen, Crafoord Foundation and from the Swedish Research Council (Vetenskapsrådet, VR) under grant 2022-03090. EvL also acknowledges support by eSSENCE, a strategic research area for e-Science, grant number eSSENCE@LU 9:1.
HURS acknowledges funding from the European Research Council (ERC) under the European
 Union’s Horizon 2020 research and innovation programme (Grant agreement No.\ 854843-FASTCORR).
The computations were enabled by resources provided by the National Academic Infrastructure for Supercomputing in Sweden (NAISS) and the Swedish National Infrastructure for Computing (SNIC)
through the projects
LU 2023/2-37, 
LU 2022/2-32, 
SNIC 2022/23-304, 
SNIC 2022/21-15, 
SNIC 2022/13-9, 
SNIC 2022/6-113, 
and 
SNIC 2022/1-18, 
at Lunarc, PDC, NSC and CSC partially funded by the Swedish Research Council through grant agreements no.\ 2022-06725 and no.\ 2018-05973.

\clearpage

\appendix 

\section{Conventions for two-particle quantities}

\label{app:conventions}

As discussed in the main text, the two-particle Green's functions of the impurity problem are a necessary ingredient for the dual Bethe-Salpeter equation. They are obtained from the impurity solver, in this case TRIQS/cthyb or w2dynamics. There are small differences in the definition of these objects between the two codes, which are essential for a correct implementation. This appendix provides their relation. 

\subsection{Two-particle Green's function in the particle-hole channel}

The four-point response function in TRIQS/cthyb for the particle-hole channel (PH) is
\begin{multline}
  g^{(4)ph}_{abcd}(\omega, \nu, \nu') =
  \frac{1}{\beta} \iiiint_0^\beta d\tau_1 d\tau_2 d\tau_3 d\tau_4
  \\
  \exp \left[ i\omega(\tau_2 - \tau_3) + i\nu(\tau_2 - \tau_1) + i\nu'(\tau_4 - \tau_3) \right]
  \\
  \langle \mathcal{T} c^\dagger_a(\tau_1) c_b(\tau_2) c^\dagger_c(\tau_3) c_d(\tau_4) \rangle
  \label{eq:g4_triqs}
\end{multline}
In the w2dynamics documentation, the four-point response function for the particle-hole channel (PH) is defined as
\begin{multline}
  \tilde{g}^{(4)ph}_{abcd}(\omega, \nu, \nu') =
  \iiiint_0^\beta d\tau_1 d\tau_2 d\tau_3 d\tau_4
  \\
  \exp \left[ i\omega(\tau_1 - \tau_4) + i\nu(\tau_1 - \tau_2) + i\nu'(\tau_3 - \tau_4) \right]
  \\
  \langle \mathcal{T} c_a(\tau_1) c^\dagger_b(\tau_2) c_c(\tau_3) c^\dagger_d(\tau_4) \rangle
  \\ =
  \iiiint_0^\beta d\tau_1 d\tau_2 d\tau_3 d\tau_4
  \\
  \exp \left[ i\omega(\tau_2 - \tau_3) + i\nu(\tau_2 - \tau_1) + i\nu'(\tau_4 - \tau_3) \right]
  \\
  \langle \mathcal{T} c^\dagger_b(\tau_1) c_a(\tau_2) c^\dagger_d(\tau_3) c_c(\tau_4) \rangle
  =
  \beta g^{(4)ph}_{badc}(\omega, \nu, \nu')
  \label{eq:g4_w2dyn}
\end{multline}
Note that the bosonic frequency convention in W2Dynamics is controlled by setting the parameter \verb|WormPHConvention=0| to get the $i\omega(\tau_1 - \tau_4)$ factor (setting \verb|WormPHConvention=1| gives $i\omega(\tau_2 - \tau_3)$ and is not directly compatible with TRIQS/cthyb+TPRF).

In numerical comparisons, we have found that the actual output data of both solvers are related as
\begin{equation}
  g^{(4)ph}_{abcd}(\omega, \nu, \nu') = \beta \tilde{g}^{(4)ph}_{badc}(\omega, \nu, \nu'),
\end{equation}
which differs from Eq.~\eqref{eq:g4_w2dyn} by a factor $\beta^2$. This is the relation used in our implementation. It should be noted that factors of $\beta$ are not always written explicitly in parts of the literature, which can make it harder to translate equations from one notation to another. 

\subsection{Two-particle Green's function in the particle-hole channel with $\tau_3 = \tau_4$}

The two-particle Green's function with $\tau_3 = \tau_4$ following the TRIQS/cthyb PH format from Eq.~(\ref{eq:g4_triqs}) is
\begin{multline}
  g^{(3)ph}_{abcd}(\omega, \nu)
  =
  \iiint_0^\beta d\tau_1 d\tau_2 d\tau_3
  \\ \times
  \exp[ i\omega ( \tau_2 - \tau_3 ) + i\nu (\tau_2 - \tau_1) ]
  \\ \times
  \langle \mathcal{T} c^\dagger_a(\tau_1) c_b(\tau_2) c^\dagger_c(\tau_3) c_d(\tau_3) \rangle  .
\end{multline}
The sampled quantity as defined from W2Dynamics is
\begin{multline}
  \tilde{g}^{(3)ph}_{abcd}(\omega, \nu)
  =
  \iiint_0^\beta d\tau_1 d\tau_2 d\tau_3
  \\ \times
  \exp[ i\omega ( \tau_2 - \tau_3 ) + i\nu (\tau_1 - \tau_2) ]
  \\ \times
  \langle \mathcal{T} c_a(\tau_1) c^\dagger_b(\tau_2) c_c(\tau_3) c^\dagger_d(\tau_3) \rangle
  \\ =
  \iiint_0^\beta d\tau_1 d\tau_2 d\tau_3
  \exp[ i\omega ( \tau_1 - \tau_3 ) + i\nu (\tau_2 - \tau_1) ]
  \\ \times
  \langle \mathcal{T} c^\dagger_b(\tau_1) c_a(\tau_2) c^\dagger_d(\tau_3) c_c(\tau_3) \rangle  
  \\ =
  \iiint_0^\beta d\tau_1 d\tau_2 d\tau_3
  \exp[ i\omega ( \tau_2 - \tau_3 ) + i(\nu - \omega)(\tau_2 - \tau_1) ]
  \\ \times
  \langle \mathcal{T} c^\dagger_b(\tau_1) c_a(\tau_2) c^\dagger_d(\tau_3) c_c(\tau_3) \rangle
  =
  g^{(3)ph}_{badc}(\omega, \nu - \omega) . 
\end{multline}
Note that the bosonic frequency convention produces a shift in the fermionic frequency. The implementation is not sensitive to the \verb|WormPHConvention| parameter. Hence both orbital label pair-permutations and a bosonic frequency shift are required to transform from W2Dynamics to TRIQS format
\begin{equation}
  g^{(3)ph}_{abcd}(\omega, \nu)
  =
  \tilde{g}^{(3)ph}_{badc}(\omega, \nu + \omega).
\end{equation}
The fermionic frequency shift, which is somewhat inconvenient, can be avoided by using time-reversal symmetry, as in Eqs.~\eqref{eq:L:symmetry} and \eqref{eq:L:symmetry:real},
\begin{align} 
 g^{(3)ph}_{badc}(\omega, -\nu)^\ast = g^{(3)ph}_{abcd}(\omega, \nu-\omega) = \tilde{g}^{(3)ph}_{badc}(\omega, \nu), \notag
\\
 g^{(3)ph}_{badc}(-\omega, \nu) \overset{\text{real }\hat{H}}{=} g^{(3)ph}_{abcd}(\omega, \nu-\omega) = \tilde{g}^{(3)ph}_{badc}(\omega, \nu). \notag
\end{align}

\subsection{Symmetry relation $L$}
\label{app:L:symmetry}

Performing complex conjugation on the relevant expectation value gives
\begin{align}
   g^{(3)}_{abcd}(\tau_1,\tau_2,\tau_3)^\ast=&\langle \mathcal{T} c^\dagger_a(\tau_1) c_b(\tau_2) c^\dagger_c(\tau_3) c_d(\tau_3) \rangle^\ast \\
   =& \langle \mathcal{T}  c^\dagger_d(-\tau_3) c_c(-\tau_3) c^\dagger_b(-\tau_2) c_a(-\tau_1) \rangle \notag \\
   =& \langle \mathcal{T}  c^\dagger_b(-\tau_2) c_a(-\tau_1) c^\dagger_d(-\tau_3) c_c(-\tau_3)  \rangle \notag \\
   =& g^{(3)}_{badc}(-\tau_2,-\tau_1,-\tau_3)
\end{align}
Now, applying a time shift $\tau_1+\tau_2$ to all arguments and using time-translation symmetry gives
\begin{equation}
 g^{(3)}_{abcd}(\tau_1,\tau_2,\tau_3)^\ast = g^{(3)}_{badc}(-\tau_2,-\tau_1,-\tau_3)
\end{equation}
For the Fourier transform to frequencies, this means
\begin{multline}
 \left[ g^{(3)}_{abcd}(\omega,\nu) \right]^*
 \\ =
 \iiint_0^\beta \!\!\! d^3\tau_i \,
 e^{-i\omega (\tau_2 - \tau_3) - i\nu (\tau_2 - \tau_1)}
 g^{(3)}_{badc}(-\tau_2, -\tau_1, -\tau_3)
 \\
 = \iiint_{-\beta}^0 \!\!\! d^3\tau_i' \,
 e^{ i\omega (\tau_1' - \tau_3') + i\nu (\tau_1' - \tau_2') }
 g^{(3)}_{badc}(\tau_1', \tau_2', \tau_3')
 \\
 = \iiint_0^{\beta} \!\!\! d^3\tau_i'
 e^{ i\omega (\tau_2' - \tau_3') + i[-\nu - \omega] (\tau_2' - \tau_1') }
 g^{(3)}_{badc}(\tau_1', \tau_2', \tau_3')
 \\
 = g^{(3)}_{badc}(\omega,-\nu-\omega)
\end{multline}
where the change of variables $\tau_1' = - \tau_2$, $\tau_2' = -\tau_1$, $\tau_3' = -\tau_3$ and the (anti)-periodicity of the integrand has been used.
Writing the symmetry relation as
\begin{equation}
  \left[ g^{(3)}_{abcd}(\omega,\nu - \omega/2) \right]^* = g^{(3)}_{badc}(\omega,-\nu-\omega/2)
  \, ,
  \label{eq:g3sym}
\end{equation}
clearly shows the role of $\omega/2$ as the center of the fermionic frequency structure. On the other hand, the measured box typically takes $\nu \in [-N_\nu,N_\nu]$. For $\omega > 0$, we can use the symmetry relation to map some frequencies outside of the frequency box to those inside of the frequency box. A similar relationship for density fluctuations in the single-orbital case was derived in Appendix F of Ref.~\onlinecite{vanLoon14}, albeit with other notational conventions.

For real Hamiltonians, we can derive a further symmetry relation, since all relevant imaginary time expectation values are real. Thus, $g^{(3)}_{abcd}(\tau_1,\tau_2,\tau_3)=g^{(3)}_{badc}(-\tau_2,-\tau_1,-\tau_3)$ and
\begin{align}
 &g^{(3)}_{abcd}(\omega,\nu-\omega/2) \notag \\
 &= \iiint_0^\beta d^3 \tau_1 e^{i\omega(\tau_2-\tau_3) + i(\nu-\omega/2)(\tau_2-\tau_1)} g^{(3)}_{badc}(-\tau_2,-\tau_1,-\tau_3)
 \notag \\
&= \iiint_0^\beta d^3 \tau_1 e^{-i\omega(\tau_1'-\tau_3') + i(\nu-\omega/2)(-\tau_1'+\tau_2')} g^{(3)}_{badc}(\tau_1',\tau_2',\tau_3')
 \notag \\
&= \iiint_0^\beta d^3 \tau_1 e^{-i\omega(\tau_2'-\tau_3') +i\omega(\tau_2'-\tau_1')+ i(\nu-\omega/2)(\tau_2'-\tau_1')} g^{(3)}_{badc}(\tau_1',\tau_2',\tau_3')
 \notag \\
 &= \iiint_0^\beta d^3 \tau_1 e^{-i\omega(\tau_2'-\tau_3') + i(\nu+\omega/2)(\tau_2'-\tau_1')} g^{(3)}_{badc}(\tau_1',\tau_2',\tau_3')
 \notag \\
 &= g^{(3)}_{badc}(-\omega,\nu+\omega/2) \,\,\,\,\, \text{  for real $\hat{H}$.} \label{eq:g3sym:real}
\end{align}
Here, we substituted $\tau_1'=-\tau_2$, $\tau_2'=-\tau_1$ and $\tau_3'=-\tau_3$. 

From Eqs.\ \eqref{eq:X0} and \eqref{eq:L:amputate:2} it is possible to show that the triangular vertex $L$ obeys the same conjugate symmetry as $g^{(3)}$ in Eqs.\ \eqref{eq:g3sym} and \eqref{eq:g3sym:real}. This completes the proof of Eqs.~\eqref{eq:L:symmetry} and \eqref{eq:L:symmetry:real}.

\bibliography{references}

\begin{thebibliography}{75}%
\makeatletter
\providecommand \@ifxundefined [1]{%
 \@ifx{#1\undefined}
}%
\providecommand \@ifnum [1]{%
 \ifnum #1\expandafter \@firstoftwo
 \else \expandafter \@secondoftwo
 \fi
}%
\providecommand \@ifx [1]{%
 \ifx #1\expandafter \@firstoftwo
 \else \expandafter \@secondoftwo
 \fi
}%
\providecommand \natexlab [1]{#1}%
\providecommand \enquote  [1]{``#1''}%
\providecommand \bibnamefont  [1]{#1}%
\providecommand \bibfnamefont [1]{#1}%
\providecommand \citenamefont [1]{#1}%
\providecommand \href@noop [0]{\@secondoftwo}%
\providecommand \href [0]{\begingroup \@sanitize@url \@href}%
\providecommand \@href[1]{\@@startlink{#1}\@@href}%
\providecommand \@@href[1]{\endgroup#1\@@endlink}%
\providecommand \@sanitize@url [0]{\catcode `\\12\catcode `\$12\catcode
  `\&12\catcode `\#12\catcode `\^12\catcode `\_12\catcode `\%12\relax}%
\providecommand \@@startlink[1]{}%
\providecommand \@@endlink[0]{}%
\providecommand \url  [0]{\begingroup\@sanitize@url \@url }%
\providecommand \@url [1]{\endgroup\@href {#1}{\urlprefix }}%
\providecommand \urlprefix  [0]{URL }%
\providecommand \Eprint [0]{\href }%
\providecommand \doibase [0]{https://doi.org/}%
\providecommand \selectlanguage [0]{\@gobble}%
\providecommand \bibinfo  [0]{\@secondoftwo}%
\providecommand \bibfield  [0]{\@secondoftwo}%
\providecommand \translation [1]{[#1]}%
\providecommand \BibitemOpen [0]{}%
\providecommand \bibitemStop [0]{}%
\providecommand \bibitemNoStop [0]{.\EOS\space}%
\providecommand \EOS [0]{\spacefactor3000\relax}%
\providecommand \BibitemShut  [1]{\csname bibitem#1\endcsname}%
\let\auto@bib@innerbib\@empty
\bibitem [{\citenamefont {Imada}\ \emph {et~al.}(1998)\citenamefont {Imada},
  \citenamefont {Fujimori},\ and\ \citenamefont {Tokura}}]{Imada98}%
  \BibitemOpen
  \bibfield  {author} {\bibinfo {author} {\bibfnamefont {M.}~\bibnamefont
  {Imada}}, \bibinfo {author} {\bibfnamefont {A.}~\bibnamefont {Fujimori}},\
  and\ \bibinfo {author} {\bibfnamefont {Y.}~\bibnamefont {Tokura}},\
  }\bibfield  {title} {\bibinfo {title} {Metal-insulator transitions},\ }\href
  {https://doi.org/10.1103/RevModPhys.70.1039} {\bibfield  {journal} {\bibinfo
  {journal} {Rev. Mod. Phys.}\ }\textbf {\bibinfo {volume} {70}},\ \bibinfo
  {pages} {1039} (\bibinfo {year} {1998})}\BibitemShut {NoStop}%
\bibitem [{\citenamefont {Tamai}\ \emph {et~al.}(2019)\citenamefont {Tamai},
  \citenamefont {Zingl}, \citenamefont {Rozbicki}, \citenamefont {Cappelli},
  \citenamefont {Ricc\`o}, \citenamefont {de~la Torre}, \citenamefont
  {McKeown~Walker}, \citenamefont {Bruno}, \citenamefont {King}, \citenamefont
  {Meevasana}, \citenamefont {Shi}, \citenamefont
  {Radovi\ifmmode~\acute{c}\else \'{c}\fi{}}, \citenamefont {Plumb},
  \citenamefont {Gibbs}, \citenamefont {Mackenzie}, \citenamefont {Berthod},
  \citenamefont {Strand}, \citenamefont {Kim}, \citenamefont {Georges},\ and\
  \citenamefont {Baumberger}}]{Tamai19}%
  \BibitemOpen
  \bibfield  {author} {\bibinfo {author} {\bibfnamefont {A.}~\bibnamefont
  {Tamai}}, \bibinfo {author} {\bibfnamefont {M.}~\bibnamefont {Zingl}},
  \bibinfo {author} {\bibfnamefont {E.}~\bibnamefont {Rozbicki}}, \bibinfo
  {author} {\bibfnamefont {E.}~\bibnamefont {Cappelli}}, \bibinfo {author}
  {\bibfnamefont {S.}~\bibnamefont {Ricc\`o}}, \bibinfo {author} {\bibfnamefont
  {A.}~\bibnamefont {de~la Torre}}, \bibinfo {author} {\bibfnamefont
  {S.}~\bibnamefont {McKeown~Walker}}, \bibinfo {author} {\bibfnamefont
  {F.~Y.}\ \bibnamefont {Bruno}}, \bibinfo {author} {\bibfnamefont {P.~D.~C.}\
  \bibnamefont {King}}, \bibinfo {author} {\bibfnamefont {W.}~\bibnamefont
  {Meevasana}}, \bibinfo {author} {\bibfnamefont {M.}~\bibnamefont {Shi}},
  \bibinfo {author} {\bibfnamefont {M.}~\bibnamefont
  {Radovi\ifmmode~\acute{c}\else \'{c}\fi{}}}, \bibinfo {author} {\bibfnamefont
  {N.~C.}\ \bibnamefont {Plumb}}, \bibinfo {author} {\bibfnamefont {A.~S.}\
  \bibnamefont {Gibbs}}, \bibinfo {author} {\bibfnamefont {A.~P.}\ \bibnamefont
  {Mackenzie}}, \bibinfo {author} {\bibfnamefont {C.}~\bibnamefont {Berthod}},
  \bibinfo {author} {\bibfnamefont {H.~U.~R.}\ \bibnamefont {Strand}}, \bibinfo
  {author} {\bibfnamefont {M.}~\bibnamefont {Kim}}, \bibinfo {author}
  {\bibfnamefont {A.}~\bibnamefont {Georges}},\ and\ \bibinfo {author}
  {\bibfnamefont {F.}~\bibnamefont {Baumberger}},\ }\bibfield  {title}
  {\bibinfo {title} {High-resolution photoemission on {Sr$_2$RuO$_4$} reveals
  correlation-enhanced effective spin-orbit coupling and dominantly local
  self-energies},\ }\href {https://doi.org/10.1103/PhysRevX.9.021048}
  {\bibfield  {journal} {\bibinfo  {journal} {Phys. Rev. X}\ }\textbf {\bibinfo
  {volume} {9}},\ \bibinfo {pages} {021048} (\bibinfo {year}
  {2019})}\BibitemShut {NoStop}%
\bibitem [{\citenamefont {Coldea}\ and\ \citenamefont
  {Watson}(2018)}]{Coldea18}%
  \BibitemOpen
  \bibfield  {author} {\bibinfo {author} {\bibfnamefont {A.~I.}\ \bibnamefont
  {Coldea}}\ and\ \bibinfo {author} {\bibfnamefont {M.~D.}\ \bibnamefont
  {Watson}},\ }\bibfield  {title} {\bibinfo {title} {The key ingredients of the
  electronic structure of {FeSe}},\ }\href
  {https://doi.org/10.1146/annurev-conmatphys-033117-054137} {\bibfield
  {journal} {\bibinfo  {journal} {Annual Review of Condensed Matter Physics}\
  }\textbf {\bibinfo {volume} {9}},\ \bibinfo {pages} {125} (\bibinfo {year}
  {2018})},\ \Eprint
  {https://arxiv.org/abs/https://doi.org/10.1146/annurev-conmatphys-033117-054137}
  {https://doi.org/10.1146/annurev-conmatphys-033117-054137} \BibitemShut
  {NoStop}%
\bibitem [{\citenamefont {Phillips}\ \emph {et~al.}(2022)\citenamefont
  {Phillips}, \citenamefont {Hussey},\ and\ \citenamefont
  {Abbamonte}}]{Phillips22}%
  \BibitemOpen
  \bibfield  {author} {\bibinfo {author} {\bibfnamefont {P.~W.}\ \bibnamefont
  {Phillips}}, \bibinfo {author} {\bibfnamefont {N.~E.}\ \bibnamefont
  {Hussey}},\ and\ \bibinfo {author} {\bibfnamefont {P.}~\bibnamefont
  {Abbamonte}},\ }\bibfield  {title} {\bibinfo {title} {Stranger than metals},\
  }\href {https://doi.org/10.1126/science.abh4273} {\bibfield  {journal}
  {\bibinfo  {journal} {Science}\ }\textbf {\bibinfo {volume} {377}},\ \bibinfo
  {pages} {eabh4273} (\bibinfo {year} {2022})}\BibitemShut {NoStop}%
\bibitem [{\citenamefont {Metzner}\ and\ \citenamefont
  {Vollhardt}(1989)}]{Metzner89}%
  \BibitemOpen
  \bibfield  {author} {\bibinfo {author} {\bibfnamefont {W.}~\bibnamefont
  {Metzner}}\ and\ \bibinfo {author} {\bibfnamefont {D.}~\bibnamefont
  {Vollhardt}},\ }\bibfield  {title} {\bibinfo {title} {Correlated lattice
  fermions in $d=\ensuremath{\infty}$ dimensions},\ }\href
  {https://doi.org/10.1103/PhysRevLett.62.324} {\bibfield  {journal} {\bibinfo
  {journal} {Phys. Rev. Lett.}\ }\textbf {\bibinfo {volume} {62}},\ \bibinfo
  {pages} {324} (\bibinfo {year} {1989})}\BibitemShut {NoStop}%
\bibitem [{\citenamefont {Georges}\ \emph {et~al.}(1996)\citenamefont
  {Georges}, \citenamefont {Kotliar}, \citenamefont {Krauth},\ and\
  \citenamefont {Rozenberg}}]{Georges96}%
  \BibitemOpen
  \bibfield  {author} {\bibinfo {author} {\bibfnamefont {A.}~\bibnamefont
  {Georges}}, \bibinfo {author} {\bibfnamefont {G.}~\bibnamefont {Kotliar}},
  \bibinfo {author} {\bibfnamefont {W.}~\bibnamefont {Krauth}},\ and\ \bibinfo
  {author} {\bibfnamefont {M.~J.}\ \bibnamefont {Rozenberg}},\ }\bibfield
  {title} {\bibinfo {title} {Dynamical mean-field theory of strongly correlated
  fermion systems and the limit of infinite dimensions},\ }\href
  {https://doi.org/10.1103/RevModPhys.68.13} {\bibfield  {journal} {\bibinfo
  {journal} {Rev. Mod. Phys.}\ }\textbf {\bibinfo {volume} {68}},\ \bibinfo
  {pages} {13} (\bibinfo {year} {1996})}\BibitemShut {NoStop}%
\bibitem [{\citenamefont {Rohringer}\ \emph {et~al.}(2018)\citenamefont
  {Rohringer}, \citenamefont {Hafermann}, \citenamefont {Toschi}, \citenamefont
  {Katanin}, \citenamefont {Antipov}, \citenamefont {Katsnelson}, \citenamefont
  {Lichtenstein}, \citenamefont {Rubtsov},\ and\ \citenamefont
  {Held}}]{Rohringer18}%
  \BibitemOpen
  \bibfield  {author} {\bibinfo {author} {\bibfnamefont {G.}~\bibnamefont
  {Rohringer}}, \bibinfo {author} {\bibfnamefont {H.}~\bibnamefont
  {Hafermann}}, \bibinfo {author} {\bibfnamefont {A.}~\bibnamefont {Toschi}},
  \bibinfo {author} {\bibfnamefont {A.~A.}\ \bibnamefont {Katanin}}, \bibinfo
  {author} {\bibfnamefont {A.~E.}\ \bibnamefont {Antipov}}, \bibinfo {author}
  {\bibfnamefont {M.~I.}\ \bibnamefont {Katsnelson}}, \bibinfo {author}
  {\bibfnamefont {A.~I.}\ \bibnamefont {Lichtenstein}}, \bibinfo {author}
  {\bibfnamefont {A.~N.}\ \bibnamefont {Rubtsov}},\ and\ \bibinfo {author}
  {\bibfnamefont {K.}~\bibnamefont {Held}},\ }\bibfield  {title} {\bibinfo
  {title} {Diagrammatic routes to nonlocal correlations beyond dynamical mean
  field theory},\ }\href {https://doi.org/10.1103/RevModPhys.90.025003}
  {\bibfield  {journal} {\bibinfo  {journal} {Rev. Mod. Phys.}\ }\textbf
  {\bibinfo {volume} {90}},\ \bibinfo {pages} {025003} (\bibinfo {year}
  {2018})}\BibitemShut {NoStop}%
\bibitem [{\citenamefont {Qin}\ \emph {et~al.}(2022)\citenamefont {Qin},
  \citenamefont {Sch\"{a}fer}, \citenamefont {Andergassen}, \citenamefont
  {Corboz},\ and\ \citenamefont {Gull}}]{Qin22}%
  \BibitemOpen
  \bibfield  {author} {\bibinfo {author} {\bibfnamefont {M.}~\bibnamefont
  {Qin}}, \bibinfo {author} {\bibfnamefont {T.}~\bibnamefont {Sch\"{a}fer}},
  \bibinfo {author} {\bibfnamefont {S.}~\bibnamefont {Andergassen}}, \bibinfo
  {author} {\bibfnamefont {P.}~\bibnamefont {Corboz}},\ and\ \bibinfo {author}
  {\bibfnamefont {E.}~\bibnamefont {Gull}},\ }\bibfield  {title} {\bibinfo
  {title} {The {Hubbard} model: A computational perspective},\ }\href
  {https://doi.org/10.1146/annurev-conmatphys-090921-033948} {\bibfield
  {journal} {\bibinfo  {journal} {Annual Review of Condensed Matter Physics}\
  }\textbf {\bibinfo {volume} {13}},\ \bibinfo {pages} {275} (\bibinfo {year}
  {2022})},\ \Eprint
  {https://arxiv.org/abs/https://doi.org/10.1146/annurev-conmatphys-090921-033948}
  {https://doi.org/10.1146/annurev-conmatphys-090921-033948} \BibitemShut
  {NoStop}%
\bibitem [{\citenamefont {Ribic}\ \emph {et~al.}(2017)\citenamefont {Ribic},
  \citenamefont {Gunacker}, \citenamefont {Iskakov}, \citenamefont
  {Wallerberger}, \citenamefont {Rohringer}, \citenamefont {Rubtsov},
  \citenamefont {Gull},\ and\ \citenamefont {Held}}]{Ribic17}%
  \BibitemOpen
  \bibfield  {author} {\bibinfo {author} {\bibfnamefont {T.}~\bibnamefont
  {Ribic}}, \bibinfo {author} {\bibfnamefont {P.}~\bibnamefont {Gunacker}},
  \bibinfo {author} {\bibfnamefont {S.}~\bibnamefont {Iskakov}}, \bibinfo
  {author} {\bibfnamefont {M.}~\bibnamefont {Wallerberger}}, \bibinfo {author}
  {\bibfnamefont {G.}~\bibnamefont {Rohringer}}, \bibinfo {author}
  {\bibfnamefont {A.~N.}\ \bibnamefont {Rubtsov}}, \bibinfo {author}
  {\bibfnamefont {E.}~\bibnamefont {Gull}},\ and\ \bibinfo {author}
  {\bibfnamefont {K.}~\bibnamefont {Held}},\ }\bibfield  {title} {\bibinfo
  {title} {Role of three-particle vertex within dual fermion calculations},\
  }\href {https://doi.org/10.1103/PhysRevB.96.235127} {\bibfield  {journal}
  {\bibinfo  {journal} {Phys. Rev. B}\ }\textbf {\bibinfo {volume} {96}},\
  \bibinfo {pages} {235127} (\bibinfo {year} {2017})}\BibitemShut {NoStop}%
\bibitem [{\citenamefont {Kappl}\ \emph {et~al.}(2022)\citenamefont {Kappl},
  \citenamefont {Krien}, \citenamefont {Watzenböck},\ and\ \citenamefont
  {Held}}]{Kappl22}%
  \BibitemOpen
  \bibfield  {author} {\bibinfo {author} {\bibfnamefont {P.}~\bibnamefont
  {Kappl}}, \bibinfo {author} {\bibfnamefont {F.}~\bibnamefont {Krien}},
  \bibinfo {author} {\bibfnamefont {C.}~\bibnamefont {Watzenböck}},\ and\
  \bibinfo {author} {\bibfnamefont {K.}~\bibnamefont {Held}},\ }\href
  {https://doi.org/10.48550/ARXIV.2212.11877} {\bibinfo {title} {Non-linear
  responses and three-particle correlators in correlated electron systems
  exemplified by the {Anderson} impurity model}} (\bibinfo {year}
  {2022})\BibitemShut {NoStop}%
\bibitem [{\citenamefont {Enderle}(2014)}]{Enderle:2014}%
  \BibitemOpen
  \bibfield  {author} {\bibinfo {author} {\bibfnamefont {M.}~\bibnamefont
  {Enderle}},\ }\bibfield  {title} {\bibinfo {title} {Neutrons and magnetism},\
  }\href@noop {} {\bibfield  {journal} {\bibinfo  {journal} {Collection SFN}\
  }\textbf {\bibinfo {volume} {13}} (\bibinfo {year} {2014})}\BibitemShut
  {NoStop}%
\bibitem [{\citenamefont {Platzman}\ and\ \citenamefont
  {Wolff}(1973)}]{Platzman73}%
  \BibitemOpen
  \bibfield  {author} {\bibinfo {author} {\bibfnamefont {P.~M.}\ \bibnamefont
  {Platzman}}\ and\ \bibinfo {author} {\bibfnamefont {P.~A.}\ \bibnamefont
  {Wolff}},\ }\href@noop {} {\emph {\bibinfo {title} {Waves and interactions in
  solid state plasmas}}},\ Vol.~\bibinfo {volume} {13}\ (\bibinfo  {publisher}
  {Academic Press New York},\ \bibinfo {year} {1973})\BibitemShut {NoStop}%
\bibitem [{\citenamefont {Ament}\ \emph {et~al.}(2011)\citenamefont {Ament},
  \citenamefont {van Veenendaal}, \citenamefont {Devereaux}, \citenamefont
  {Hill},\ and\ \citenamefont {van~den Brink}}]{RevModPhys.83.705}%
  \BibitemOpen
  \bibfield  {author} {\bibinfo {author} {\bibfnamefont {L.~J.~P.}\
  \bibnamefont {Ament}}, \bibinfo {author} {\bibfnamefont {M.}~\bibnamefont
  {van Veenendaal}}, \bibinfo {author} {\bibfnamefont {T.~P.}\ \bibnamefont
  {Devereaux}}, \bibinfo {author} {\bibfnamefont {J.~P.}\ \bibnamefont
  {Hill}},\ and\ \bibinfo {author} {\bibfnamefont {J.}~\bibnamefont {van~den
  Brink}},\ }\bibfield  {title} {\bibinfo {title} {Resonant inelastic x-ray
  scattering studies of elementary excitations},\ }\href@noop {} {\bibfield
  {journal} {\bibinfo  {journal} {Rev. Mod. Phys.}\ }\textbf {\bibinfo {volume}
  {83}},\ \bibinfo {pages} {705} (\bibinfo {year} {2011})}\BibitemShut
  {NoStop}%
\bibitem [{\citenamefont {Kune{\v s}}(2011)}]{PhysRevB.83.085102}%
  \BibitemOpen
  \bibfield  {author} {\bibinfo {author} {\bibfnamefont {J.}~\bibnamefont
  {Kune{\v s}}},\ }\bibfield  {title} {\bibinfo {title} {Efficient treatment of
  two-particle vertices in dynamical mean-field theory},\ }\href@noop {}
  {\bibfield  {journal} {\bibinfo  {journal} {Phys. Rev. B}\ }\textbf {\bibinfo
  {volume} {83}},\ \bibinfo {pages} {085102} (\bibinfo {year}
  {2011})}\BibitemShut {NoStop}%
\bibitem [{\citenamefont {Boehnke}\ \emph {et~al.}(2011)\citenamefont
  {Boehnke}, \citenamefont {Hafermann}, \citenamefont {Ferrero}, \citenamefont
  {Lechermann},\ and\ \citenamefont {Parcollet}}]{Boehnke:2011fk}%
  \BibitemOpen
  \bibfield  {author} {\bibinfo {author} {\bibfnamefont {L.}~\bibnamefont
  {Boehnke}}, \bibinfo {author} {\bibfnamefont {H.}~\bibnamefont {Hafermann}},
  \bibinfo {author} {\bibfnamefont {M.}~\bibnamefont {Ferrero}}, \bibinfo
  {author} {\bibfnamefont {F.}~\bibnamefont {Lechermann}},\ and\ \bibinfo
  {author} {\bibfnamefont {O.}~\bibnamefont {Parcollet}},\ }\bibfield  {title}
  {\bibinfo {title} {Orthogonal polynomial representation of imaginary-time
  {Green's} functions},\ }\href@noop {} {\bibfield  {journal} {\bibinfo
  {journal} {Phys. Rev. B}\ }\textbf {\bibinfo {volume} {84}},\ \bibinfo
  {pages} {075145} (\bibinfo {year} {2011})}\BibitemShut {NoStop}%
\bibitem [{\citenamefont {Kaufmann}\ \emph {et~al.}(2017)\citenamefont
  {Kaufmann}, \citenamefont {Gunacker},\ and\ \citenamefont
  {Held}}]{Kaufmann17}%
  \BibitemOpen
  \bibfield  {author} {\bibinfo {author} {\bibfnamefont {J.}~\bibnamefont
  {Kaufmann}}, \bibinfo {author} {\bibfnamefont {P.}~\bibnamefont {Gunacker}},\
  and\ \bibinfo {author} {\bibfnamefont {K.}~\bibnamefont {Held}},\ }\bibfield
  {title} {\bibinfo {title} {Continuous-time quantum {Monte Carlo} calculation
  of multiorbital vertex asymptotics},\ }\href
  {https://doi.org/10.1103/PhysRevB.96.035114} {\bibfield  {journal} {\bibinfo
  {journal} {Phys. Rev. B}\ }\textbf {\bibinfo {volume} {96}},\ \bibinfo
  {pages} {035114} (\bibinfo {year} {2017})}\BibitemShut {NoStop}%
\bibitem [{\citenamefont {Tagliavini}\ \emph {et~al.}(2018)\citenamefont
  {Tagliavini}, \citenamefont {Hummel}, \citenamefont {Wentzell}, \citenamefont
  {Andergassen}, \citenamefont {Toschi},\ and\ \citenamefont
  {Rohringer}}]{Tagliavini18}%
  \BibitemOpen
  \bibfield  {author} {\bibinfo {author} {\bibfnamefont {A.}~\bibnamefont
  {Tagliavini}}, \bibinfo {author} {\bibfnamefont {S.}~\bibnamefont {Hummel}},
  \bibinfo {author} {\bibfnamefont {N.}~\bibnamefont {Wentzell}}, \bibinfo
  {author} {\bibfnamefont {S.}~\bibnamefont {Andergassen}}, \bibinfo {author}
  {\bibfnamefont {A.}~\bibnamefont {Toschi}},\ and\ \bibinfo {author}
  {\bibfnamefont {G.}~\bibnamefont {Rohringer}},\ }\bibfield  {title} {\bibinfo
  {title} {Efficient {Bethe-Salpeter} equation treatment in dynamical
  mean-field theory},\ }\href {https://doi.org/10.1103/PhysRevB.97.235140}
  {\bibfield  {journal} {\bibinfo  {journal} {Phys. Rev. B}\ }\textbf {\bibinfo
  {volume} {97}},\ \bibinfo {pages} {235140} (\bibinfo {year}
  {2018})}\BibitemShut {NoStop}%
\bibitem [{\citenamefont {Wentzell}\ \emph {et~al.}(2020)\citenamefont
  {Wentzell}, \citenamefont {Li}, \citenamefont {Tagliavini}, \citenamefont
  {Taranto}, \citenamefont {Rohringer}, \citenamefont {Held}, \citenamefont
  {Toschi},\ and\ \citenamefont {Andergassen}}]{Wentzell20}%
  \BibitemOpen
  \bibfield  {author} {\bibinfo {author} {\bibfnamefont {N.}~\bibnamefont
  {Wentzell}}, \bibinfo {author} {\bibfnamefont {G.}~\bibnamefont {Li}},
  \bibinfo {author} {\bibfnamefont {A.}~\bibnamefont {Tagliavini}}, \bibinfo
  {author} {\bibfnamefont {C.}~\bibnamefont {Taranto}}, \bibinfo {author}
  {\bibfnamefont {G.}~\bibnamefont {Rohringer}}, \bibinfo {author}
  {\bibfnamefont {K.}~\bibnamefont {Held}}, \bibinfo {author} {\bibfnamefont
  {A.}~\bibnamefont {Toschi}},\ and\ \bibinfo {author} {\bibfnamefont
  {S.}~\bibnamefont {Andergassen}},\ }\bibfield  {title} {\bibinfo {title}
  {High-frequency asymptotics of the vertex function: Diagrammatic
  parametrization and algorithmic implementation},\ }\href
  {https://doi.org/10.1103/PhysRevB.102.085106} {\bibfield  {journal} {\bibinfo
   {journal} {Phys. Rev. B}\ }\textbf {\bibinfo {volume} {102}},\ \bibinfo
  {pages} {085106} (\bibinfo {year} {2020})}\BibitemShut {NoStop}%
\bibitem [{\citenamefont {Katanin}(2020)}]{PhysRevB.101.035110}%
  \BibitemOpen
  \bibfield  {author} {\bibinfo {author} {\bibfnamefont {A.}~\bibnamefont
  {Katanin}},\ }\bibfield  {title} {\bibinfo {title} {Improved treatment of
  fermion-boson vertices and bethe-salpeter equations in nonlocal extensions of
  dynamical mean field theory},\ }\href
  {https://doi.org/10.1103/PhysRevB.101.035110} {\bibfield  {journal} {\bibinfo
   {journal} {Phys. Rev. B}\ }\textbf {\bibinfo {volume} {101}},\ \bibinfo
  {pages} {035110} (\bibinfo {year} {2020})}\BibitemShut {NoStop}%
\bibitem [{\citenamefont {Pruschke}\ \emph {et~al.}(1996)\citenamefont
  {Pruschke}, \citenamefont {Qin}, \citenamefont {Obermeier},\ and\
  \citenamefont {Keller}}]{Pruschke96}%
  \BibitemOpen
  \bibfield  {author} {\bibinfo {author} {\bibfnamefont {T.}~\bibnamefont
  {Pruschke}}, \bibinfo {author} {\bibfnamefont {Q.}~\bibnamefont {Qin}},
  \bibinfo {author} {\bibfnamefont {T.}~\bibnamefont {Obermeier}},\ and\
  \bibinfo {author} {\bibfnamefont {J.}~\bibnamefont {Keller}},\ }\bibfield
  {title} {\bibinfo {title} {Magnetic properties of the t - j model in the
  dynamical mean-field theory},\ }\href
  {https://doi.org/10.1088/0953-8984/8/18/009} {\bibfield  {journal} {\bibinfo
  {journal} {Journal of Physics: Condensed Matter}\ }\textbf {\bibinfo {volume}
  {8}},\ \bibinfo {pages} {3161} (\bibinfo {year} {1996})}\BibitemShut
  {NoStop}%
\bibitem [{\citenamefont {Rubtsov}\ \emph {et~al.}(2012)\citenamefont
  {Rubtsov}, \citenamefont {Katsnelson},\ and\ \citenamefont
  {Lichtenstein}}]{Rubtsov12}%
  \BibitemOpen
  \bibfield  {author} {\bibinfo {author} {\bibfnamefont {A.}~\bibnamefont
  {Rubtsov}}, \bibinfo {author} {\bibfnamefont {M.}~\bibnamefont
  {Katsnelson}},\ and\ \bibinfo {author} {\bibfnamefont {A.}~\bibnamefont
  {Lichtenstein}},\ }\bibfield  {title} {\bibinfo {title} {Dual boson approach
  to collective excitations in correlated fermionic systems},\ }\href
  {https://doi.org/https://doi.org/10.1016/j.aop.2012.01.002} {\bibfield
  {journal} {\bibinfo  {journal} {Annals of Physics}\ }\textbf {\bibinfo
  {volume} {327}},\ \bibinfo {pages} {1320} (\bibinfo {year}
  {2012})}\BibitemShut {NoStop}%
\bibitem [{\citenamefont {Hafermann}\ \emph {et~al.}(2014)\citenamefont
  {Hafermann}, \citenamefont {van Loon}, \citenamefont {Katsnelson},
  \citenamefont {Lichtenstein},\ and\ \citenamefont {Parcollet}}]{Hafermann14}%
  \BibitemOpen
  \bibfield  {author} {\bibinfo {author} {\bibfnamefont {H.}~\bibnamefont
  {Hafermann}}, \bibinfo {author} {\bibfnamefont {E.~G. C.~P.}\ \bibnamefont
  {van Loon}}, \bibinfo {author} {\bibfnamefont {M.~I.}\ \bibnamefont
  {Katsnelson}}, \bibinfo {author} {\bibfnamefont {A.~I.}\ \bibnamefont
  {Lichtenstein}},\ and\ \bibinfo {author} {\bibfnamefont {O.}~\bibnamefont
  {Parcollet}},\ }\bibfield  {title} {\bibinfo {title} {Collective charge
  excitations of strongly correlated electrons, vertex corrections, and gauge
  invariance},\ }\href {https://doi.org/10.1103/PhysRevB.90.235105} {\bibfield
  {journal} {\bibinfo  {journal} {Phys. Rev. B}\ }\textbf {\bibinfo {volume}
  {90}},\ \bibinfo {pages} {235105} (\bibinfo {year} {2014})}\BibitemShut
  {NoStop}%
\bibitem [{\citenamefont {van Loon}\ \emph
  {et~al.}(2014{\natexlab{a}})\citenamefont {van Loon}, \citenamefont
  {Lichtenstein}, \citenamefont {Katsnelson}, \citenamefont {Parcollet},\ and\
  \citenamefont {Hafermann}}]{vanLoon14}%
  \BibitemOpen
  \bibfield  {author} {\bibinfo {author} {\bibfnamefont {E.~G. C.~P.}\
  \bibnamefont {van Loon}}, \bibinfo {author} {\bibfnamefont {A.~I.}\
  \bibnamefont {Lichtenstein}}, \bibinfo {author} {\bibfnamefont {M.~I.}\
  \bibnamefont {Katsnelson}}, \bibinfo {author} {\bibfnamefont
  {O.}~\bibnamefont {Parcollet}},\ and\ \bibinfo {author} {\bibfnamefont
  {H.}~\bibnamefont {Hafermann}},\ }\bibfield  {title} {\bibinfo {title}
  {Beyond extended dynamical mean-field theory: Dual boson approach to the
  two-dimensional extended {Hubbard} model},\ }\href
  {https://doi.org/10.1103/PhysRevB.90.235135} {\bibfield  {journal} {\bibinfo
  {journal} {Phys. Rev. B}\ }\textbf {\bibinfo {volume} {90}},\ \bibinfo
  {pages} {235135} (\bibinfo {year} {2014}{\natexlab{a}})}\BibitemShut
  {NoStop}%
\bibitem [{\citenamefont {van Loon}\ \emph {et~al.}(2015)\citenamefont {van
  Loon}, \citenamefont {Hafermann}, \citenamefont {Lichtenstein},\ and\
  \citenamefont {Katsnelson}}]{vanLoon15}%
  \BibitemOpen
  \bibfield  {author} {\bibinfo {author} {\bibfnamefont {E.~G. C.~P.}\
  \bibnamefont {van Loon}}, \bibinfo {author} {\bibfnamefont {H.}~\bibnamefont
  {Hafermann}}, \bibinfo {author} {\bibfnamefont {A.~I.}\ \bibnamefont
  {Lichtenstein}},\ and\ \bibinfo {author} {\bibfnamefont {M.~I.}\ \bibnamefont
  {Katsnelson}},\ }\bibfield  {title} {\bibinfo {title} {Thermodynamic
  consistency of the charge response in dynamical mean-field based
  approaches},\ }\href {https://doi.org/10.1103/PhysRevB.92.085106} {\bibfield
  {journal} {\bibinfo  {journal} {Phys. Rev. B}\ }\textbf {\bibinfo {volume}
  {92}},\ \bibinfo {pages} {085106} (\bibinfo {year} {2015})}\BibitemShut
  {NoStop}%
\bibitem [{\citenamefont {van Loon}\ \emph {et~al.}(2016)\citenamefont {van
  Loon}, \citenamefont {Krien}, \citenamefont {Hafermann}, \citenamefont
  {Stepanov}, \citenamefont {Lichtenstein},\ and\ \citenamefont
  {Katsnelson}}]{vanLoon16}%
  \BibitemOpen
  \bibfield  {author} {\bibinfo {author} {\bibfnamefont {E.~G. C.~P.}\
  \bibnamefont {van Loon}}, \bibinfo {author} {\bibfnamefont {F.}~\bibnamefont
  {Krien}}, \bibinfo {author} {\bibfnamefont {H.}~\bibnamefont {Hafermann}},
  \bibinfo {author} {\bibfnamefont {E.~A.}\ \bibnamefont {Stepanov}}, \bibinfo
  {author} {\bibfnamefont {A.~I.}\ \bibnamefont {Lichtenstein}},\ and\ \bibinfo
  {author} {\bibfnamefont {M.~I.}\ \bibnamefont {Katsnelson}},\ }\bibfield
  {title} {\bibinfo {title} {Double occupancy in dynamical mean-field theory
  and the dual boson approach},\ }\href
  {https://doi.org/10.1103/PhysRevB.93.155162} {\bibfield  {journal} {\bibinfo
  {journal} {Phys. Rev. B}\ }\textbf {\bibinfo {volume} {93}},\ \bibinfo
  {pages} {155162} (\bibinfo {year} {2016})}\BibitemShut {NoStop}%
\bibitem [{\citenamefont {Krien}(2019)}]{Krien19}%
  \BibitemOpen
  \bibfield  {author} {\bibinfo {author} {\bibfnamefont {F.}~\bibnamefont
  {Krien}},\ }\bibfield  {title} {\bibinfo {title} {Efficient evaluation of the
  polarization function in dynamical mean-field theory},\ }\href
  {https://doi.org/10.1103/PhysRevB.99.235106} {\bibfield  {journal} {\bibinfo
  {journal} {Phys. Rev. B}\ }\textbf {\bibinfo {volume} {99}},\ \bibinfo
  {pages} {235106} (\bibinfo {year} {2019})}\BibitemShut {NoStop}%
\bibitem [{\citenamefont {van Loon}\ \emph
  {et~al.}(2014{\natexlab{b}})\citenamefont {van Loon}, \citenamefont
  {Hafermann}, \citenamefont {Lichtenstein}, \citenamefont {Rubtsov},\ and\
  \citenamefont {Katsnelson}}]{vanLoon14PRL}%
  \BibitemOpen
  \bibfield  {author} {\bibinfo {author} {\bibfnamefont {E.~G. C.~P.}\
  \bibnamefont {van Loon}}, \bibinfo {author} {\bibfnamefont {H.}~\bibnamefont
  {Hafermann}}, \bibinfo {author} {\bibfnamefont {A.~I.}\ \bibnamefont
  {Lichtenstein}}, \bibinfo {author} {\bibfnamefont {A.~N.}\ \bibnamefont
  {Rubtsov}},\ and\ \bibinfo {author} {\bibfnamefont {M.~I.}\ \bibnamefont
  {Katsnelson}},\ }\bibfield  {title} {\bibinfo {title} {Plasmons in strongly
  correlated systems: Spectral weight transfer and renormalized dispersion},\
  }\href {https://doi.org/10.1103/PhysRevLett.113.246407} {\bibfield  {journal}
  {\bibinfo  {journal} {Phys. Rev. Lett.}\ }\textbf {\bibinfo {volume} {113}},\
  \bibinfo {pages} {246407} (\bibinfo {year} {2014}{\natexlab{b}})}\BibitemShut
  {NoStop}%
\bibitem [{\citenamefont {Krien}\ \emph {et~al.}(2017)\citenamefont {Krien},
  \citenamefont {van Loon}, \citenamefont {Hafermann}, \citenamefont {Otsuki},
  \citenamefont {Katsnelson},\ and\ \citenamefont {Lichtenstein}}]{Krien17}%
  \BibitemOpen
  \bibfield  {author} {\bibinfo {author} {\bibfnamefont {F.}~\bibnamefont
  {Krien}}, \bibinfo {author} {\bibfnamefont {E.~G. C.~P.}\ \bibnamefont {van
  Loon}}, \bibinfo {author} {\bibfnamefont {H.}~\bibnamefont {Hafermann}},
  \bibinfo {author} {\bibfnamefont {J.}~\bibnamefont {Otsuki}}, \bibinfo
  {author} {\bibfnamefont {M.~I.}\ \bibnamefont {Katsnelson}},\ and\ \bibinfo
  {author} {\bibfnamefont {A.~I.}\ \bibnamefont {Lichtenstein}},\ }\bibfield
  {title} {\bibinfo {title} {Conservation in two-particle self-consistent
  extensions of dynamical mean-field theory},\ }\href
  {https://doi.org/10.1103/PhysRevB.96.075155} {\bibfield  {journal} {\bibinfo
  {journal} {Phys. Rev. B}\ }\textbf {\bibinfo {volume} {96}},\ \bibinfo
  {pages} {075155} (\bibinfo {year} {2017})}\BibitemShut {NoStop}%
\bibitem [{\citenamefont {Galler}\ \emph {et~al.}(2019)\citenamefont {Galler},
  \citenamefont {Thunström}, \citenamefont {Kaufmann}, \citenamefont {Pickem},
  \citenamefont {Tomczak},\ and\ \citenamefont {Held}}]{Galler19}%
  \BibitemOpen
  \bibfield  {author} {\bibinfo {author} {\bibfnamefont {A.}~\bibnamefont
  {Galler}}, \bibinfo {author} {\bibfnamefont {P.}~\bibnamefont {Thunström}},
  \bibinfo {author} {\bibfnamefont {J.}~\bibnamefont {Kaufmann}}, \bibinfo
  {author} {\bibfnamefont {M.}~\bibnamefont {Pickem}}, \bibinfo {author}
  {\bibfnamefont {J.~M.}\ \bibnamefont {Tomczak}},\ and\ \bibinfo {author}
  {\bibfnamefont {K.}~\bibnamefont {Held}},\ }\bibfield  {title} {\bibinfo
  {title} {The abinitiod$\gamma$a project v1.0: Non-local correlations beyond
  and susceptibilities within dynamical mean-field theory},\ }\href
  {https://doi.org/https://doi.org/10.1016/j.cpc.2019.07.012} {\bibfield
  {journal} {\bibinfo  {journal} {Computer Physics Communications}\ }\textbf
  {\bibinfo {volume} {245}},\ \bibinfo {pages} {106847} (\bibinfo {year}
  {2019})}\BibitemShut {NoStop}%
\bibitem [{\citenamefont {Strand}(2019)}]{Strand:tprf}%
  \BibitemOpen
  \bibfield  {author} {\bibinfo {author} {\bibfnamefont {H.~U.~R.}\
  \bibnamefont {Strand}},\ }\bibfield  {title} {\bibinfo {title} {Two-particle
  response function tool-box ({TPRF}) for {TRIQS}},\ }\href@noop {} {\bibfield
  {journal} {\bibinfo  {journal} {github.com/TRIQS/tprf}\ } (\bibinfo {year}
  {2019})}\BibitemShut {NoStop}%
\bibitem [{\citenamefont {Parcollet}\ \emph {et~al.}(2015)\citenamefont
  {Parcollet}, \citenamefont {Ferrero}, \citenamefont {Ayral}, \citenamefont
  {Hafermann}, \citenamefont {Krivenko}, \citenamefont {Messio},\ and\
  \citenamefont {Seth}}]{triqs}%
  \BibitemOpen
  \bibfield  {author} {\bibinfo {author} {\bibfnamefont {O.}~\bibnamefont
  {Parcollet}}, \bibinfo {author} {\bibfnamefont {M.}~\bibnamefont {Ferrero}},
  \bibinfo {author} {\bibfnamefont {T.}~\bibnamefont {Ayral}}, \bibinfo
  {author} {\bibfnamefont {H.}~\bibnamefont {Hafermann}}, \bibinfo {author}
  {\bibfnamefont {I.}~\bibnamefont {Krivenko}}, \bibinfo {author}
  {\bibfnamefont {L.}~\bibnamefont {Messio}},\ and\ \bibinfo {author}
  {\bibfnamefont {P.}~\bibnamefont {Seth}},\ }\bibfield  {title} {\bibinfo
  {title} {Triqs: A toolbox for research on interacting quantum systems},\
  }\href {https://doi.org/http://dx.doi.org/10.1016/j.cpc.2015.04.023}
  {\bibfield  {journal} {\bibinfo  {journal} {Computer Physics Communications}\
  }\textbf {\bibinfo {volume} {196}},\ \bibinfo {pages} {398 } (\bibinfo {year}
  {2015})}\BibitemShut {NoStop}%
\bibitem [{\citenamefont {Pavarini}\ \emph {et~al.}(2014)\citenamefont
  {Pavarini}, \citenamefont {Koch}, \citenamefont {Vollhardt},\ and\
  \citenamefont {Lichtenstein}}]{Pavarini2014dmft}%
  \BibitemOpen
  \bibfield  {author} {\bibinfo {author} {\bibfnamefont {E.}~\bibnamefont
  {Pavarini}}, \bibinfo {author} {\bibfnamefont {E.}~\bibnamefont {Koch}},
  \bibinfo {author} {\bibfnamefont {D.}~\bibnamefont {Vollhardt}},\ and\
  \bibinfo {author} {\bibfnamefont {A.}~\bibnamefont {Lichtenstein}},\
  }\href@noop {} {\emph {\bibinfo {title} {{DMFT} at 25: Infinite dimensions:
  Lecture notes of the autumn school on correlated electrons 2014}}},\
  Vol.~\bibinfo {volume} {4}\ (\bibinfo  {publisher} {Forschungszentrum
  J{\"u}lich},\ \bibinfo {year} {2014})\BibitemShut {NoStop}%
\bibitem [{\citenamefont {Boehnke}(2015)}]{BoehnkeThesis}%
  \BibitemOpen
  \bibfield  {author} {\bibinfo {author} {\bibfnamefont {L.~V.}\ \bibnamefont
  {Boehnke}},\ }\emph {\bibinfo {title} {Susceptibilities in materials with
  multiple strongly correlated orbitals}},\ \href@noop {} {Ph.D. thesis},\
  \bibinfo  {school} {Staats-und Universit{\"a}tsbibliothek Hamburg Carl von
  Ossietzky} (\bibinfo {year} {2015})\BibitemShut {NoStop}%
\bibitem [{\citenamefont {Krien}(2018)}]{KrienThesis}%
  \BibitemOpen
  \bibfield  {author} {\bibinfo {author} {\bibfnamefont {F.}~\bibnamefont
  {Krien}},\ }\emph {\bibinfo {title} {Conserving dynamical mean-field
  approaches to strongly correlated systems}},\ \href@noop {} {Ph.D. thesis},\
  \bibinfo  {school} {Staats-und Universit{\"a}tsbibliothek Hamburg Carl von
  Ossietzky} (\bibinfo {year} {2018})\BibitemShut {NoStop}%
\bibitem [{\citenamefont {Strand}\ \emph {et~al.}(2019)\citenamefont {Strand},
  \citenamefont {Zingl}, \citenamefont {Wentzell}, \citenamefont {Parcollet},\
  and\ \citenamefont {Georges}}]{Strand19}%
  \BibitemOpen
  \bibfield  {author} {\bibinfo {author} {\bibfnamefont {H.~U.~R.}\
  \bibnamefont {Strand}}, \bibinfo {author} {\bibfnamefont {M.}~\bibnamefont
  {Zingl}}, \bibinfo {author} {\bibfnamefont {N.}~\bibnamefont {Wentzell}},
  \bibinfo {author} {\bibfnamefont {O.}~\bibnamefont {Parcollet}},\ and\
  \bibinfo {author} {\bibfnamefont {A.}~\bibnamefont {Georges}},\ }\bibfield
  {title} {\bibinfo {title} {Magnetic response of
  {${\mathrm{Sr}}_{2}{\mathrm{RuO}}_{4}$}: Quasi-local spin fluctuations due to
  {Hund's} coupling},\ }\href {https://doi.org/10.1103/PhysRevB.100.125120}
  {\bibfield  {journal} {\bibinfo  {journal} {Phys. Rev. B}\ }\textbf {\bibinfo
  {volume} {100}},\ \bibinfo {pages} {125120} (\bibinfo {year}
  {2019})}\BibitemShut {NoStop}%
\bibitem [{\citenamefont {Sch\"afer}\ \emph {et~al.}(2013)\citenamefont
  {Sch\"afer}, \citenamefont {Rohringer}, \citenamefont {Gunnarsson},
  \citenamefont {Ciuchi}, \citenamefont {Sangiovanni},\ and\ \citenamefont
  {Toschi}}]{Schafer13}%
  \BibitemOpen
  \bibfield  {author} {\bibinfo {author} {\bibfnamefont {T.}~\bibnamefont
  {Sch\"afer}}, \bibinfo {author} {\bibfnamefont {G.}~\bibnamefont
  {Rohringer}}, \bibinfo {author} {\bibfnamefont {O.}~\bibnamefont
  {Gunnarsson}}, \bibinfo {author} {\bibfnamefont {S.}~\bibnamefont {Ciuchi}},
  \bibinfo {author} {\bibfnamefont {G.}~\bibnamefont {Sangiovanni}},\ and\
  \bibinfo {author} {\bibfnamefont {A.}~\bibnamefont {Toschi}},\ }\bibfield
  {title} {\bibinfo {title} {Divergent precursors of the {Mott-Hubbard}
  transition at the two-particle level},\ }\href
  {https://doi.org/10.1103/PhysRevLett.110.246405} {\bibfield  {journal}
  {\bibinfo  {journal} {Phys. Rev. Lett.}\ }\textbf {\bibinfo {volume} {110}},\
  \bibinfo {pages} {246405} (\bibinfo {year} {2013})}\BibitemShut {NoStop}%
\bibitem [{\citenamefont {Chalupa}\ \emph {et~al.}(2018)\citenamefont
  {Chalupa}, \citenamefont {Gunacker}, \citenamefont {Sch\"afer}, \citenamefont
  {Held},\ and\ \citenamefont {Toschi}}]{Chalupa18}%
  \BibitemOpen
  \bibfield  {author} {\bibinfo {author} {\bibfnamefont {P.}~\bibnamefont
  {Chalupa}}, \bibinfo {author} {\bibfnamefont {P.}~\bibnamefont {Gunacker}},
  \bibinfo {author} {\bibfnamefont {T.}~\bibnamefont {Sch\"afer}}, \bibinfo
  {author} {\bibfnamefont {K.}~\bibnamefont {Held}},\ and\ \bibinfo {author}
  {\bibfnamefont {A.}~\bibnamefont {Toschi}},\ }\bibfield  {title} {\bibinfo
  {title} {Divergences of the irreducible vertex functions in correlated
  metallic systems: Insights from the {Anderson} impurity model},\ }\href
  {https://doi.org/10.1103/PhysRevB.97.245136} {\bibfield  {journal} {\bibinfo
  {journal} {Phys. Rev. B}\ }\textbf {\bibinfo {volume} {97}},\ \bibinfo
  {pages} {245136} (\bibinfo {year} {2018})}\BibitemShut {NoStop}%
\bibitem [{\citenamefont {Reitner}\ \emph {et~al.}(2020)\citenamefont
  {Reitner}, \citenamefont {Chalupa}, \citenamefont {Del~Re}, \citenamefont
  {Springer}, \citenamefont {Ciuchi}, \citenamefont {Sangiovanni},\ and\
  \citenamefont {Toschi}}]{Reitner20}%
  \BibitemOpen
  \bibfield  {author} {\bibinfo {author} {\bibfnamefont {M.}~\bibnamefont
  {Reitner}}, \bibinfo {author} {\bibfnamefont {P.}~\bibnamefont {Chalupa}},
  \bibinfo {author} {\bibfnamefont {L.}~\bibnamefont {Del~Re}}, \bibinfo
  {author} {\bibfnamefont {D.}~\bibnamefont {Springer}}, \bibinfo {author}
  {\bibfnamefont {S.}~\bibnamefont {Ciuchi}}, \bibinfo {author} {\bibfnamefont
  {G.}~\bibnamefont {Sangiovanni}},\ and\ \bibinfo {author} {\bibfnamefont
  {A.}~\bibnamefont {Toschi}},\ }\bibfield  {title} {\bibinfo {title}
  {Attractive effect of a strong electronic repulsion: The physics of vertex
  divergences},\ }\href {https://doi.org/10.1103/PhysRevLett.125.196403}
  {\bibfield  {journal} {\bibinfo  {journal} {Phys. Rev. Lett.}\ }\textbf
  {\bibinfo {volume} {125}},\ \bibinfo {pages} {196403} (\bibinfo {year}
  {2020})}\BibitemShut {NoStop}%
\bibitem [{\citenamefont {van Loon}\ \emph {et~al.}(2020)\citenamefont {van
  Loon}, \citenamefont {Krien},\ and\ \citenamefont {Katanin}}]{vanLoon20}%
  \BibitemOpen
  \bibfield  {author} {\bibinfo {author} {\bibfnamefont {E.~G. C.~P.}\
  \bibnamefont {van Loon}}, \bibinfo {author} {\bibfnamefont {F.}~\bibnamefont
  {Krien}},\ and\ \bibinfo {author} {\bibfnamefont {A.~A.}\ \bibnamefont
  {Katanin}},\ }\bibfield  {title} {\bibinfo {title} {{Bethe-Salpeter} equation
  at the critical end point of the {Mott} transition},\ }\href
  {https://doi.org/10.1103/PhysRevLett.125.136402} {\bibfield  {journal}
  {\bibinfo  {journal} {Phys. Rev. Lett.}\ }\textbf {\bibinfo {volume} {125}},\
  \bibinfo {pages} {136402} (\bibinfo {year} {2020})}\BibitemShut {NoStop}%
\bibitem [{\citenamefont {van Loon}\ \emph {et~al.}(2018)\citenamefont {van
  Loon}, \citenamefont {Krien}, \citenamefont {Hafermann}, \citenamefont
  {Lichtenstein},\ and\ \citenamefont {Katsnelson}}]{vanLoon18fermionboson}%
  \BibitemOpen
  \bibfield  {author} {\bibinfo {author} {\bibfnamefont {E.~G. C.~P.}\
  \bibnamefont {van Loon}}, \bibinfo {author} {\bibfnamefont {F.}~\bibnamefont
  {Krien}}, \bibinfo {author} {\bibfnamefont {H.}~\bibnamefont {Hafermann}},
  \bibinfo {author} {\bibfnamefont {A.~I.}\ \bibnamefont {Lichtenstein}},\ and\
  \bibinfo {author} {\bibfnamefont {M.~I.}\ \bibnamefont {Katsnelson}},\
  }\bibfield  {title} {\bibinfo {title} {Fermion-boson vertex within dynamical
  mean-field theory},\ }\href {https://doi.org/10.1103/PhysRevB.98.205148}
  {\bibfield  {journal} {\bibinfo  {journal} {Phys. Rev. B}\ }\textbf {\bibinfo
  {volume} {98}},\ \bibinfo {pages} {205148} (\bibinfo {year}
  {2018})}\BibitemShut {NoStop}%
\bibitem [{\citenamefont {Hafermann}(2010)}]{Hafermannphd}%
  \BibitemOpen
  \bibfield  {author} {\bibinfo {author} {\bibfnamefont {H.}~\bibnamefont
  {Hafermann}},\ }\emph {\bibinfo {title} {Numerical Approaches to Spatial
  Correlations in Strongly Interacting Fermion Systems}},\ \href@noop {} {Ph.D.
  thesis},\ \bibinfo  {school} {University of Hamburg} (\bibinfo {year}
  {2010})\BibitemShut {NoStop}%
\bibitem [{\citenamefont {Kaltak}(2015)}]{Kaltakphd}%
  \BibitemOpen
  \bibfield  {author} {\bibinfo {author} {\bibfnamefont {M.}~\bibnamefont
  {Kaltak}},\ }\emph {\bibinfo {title} {Merging $GW$ with DMFT}},\ \href
  {http://othes.univie.ac.at/38099/} {Ph.D. thesis},\ \bibinfo  {school}
  {Universit\"at Wien} (\bibinfo {year} {2015})\BibitemShut {NoStop}%
\bibitem [{\citenamefont {Seth}\ \emph {et~al.}(2016)\citenamefont {Seth},
  \citenamefont {Krivenko}, \citenamefont {Ferrero},\ and\ \citenamefont
  {Parcollet}}]{cthyb}%
  \BibitemOpen
  \bibfield  {author} {\bibinfo {author} {\bibfnamefont {P.}~\bibnamefont
  {Seth}}, \bibinfo {author} {\bibfnamefont {I.}~\bibnamefont {Krivenko}},
  \bibinfo {author} {\bibfnamefont {M.}~\bibnamefont {Ferrero}},\ and\ \bibinfo
  {author} {\bibfnamefont {O.}~\bibnamefont {Parcollet}},\ }\bibfield  {title}
  {\bibinfo {title} {{TRIQS/CTHYB}: A continuous-time quantum {Monte Carlo}
  hybridisation expansion solver for quantum impurity problems},\ }\href
  {https://doi.org/https://doi.org/10.1016/j.cpc.2015.10.023} {\bibfield
  {journal} {\bibinfo  {journal} {Computer Physics Communications}\ }\textbf
  {\bibinfo {volume} {200}},\ \bibinfo {pages} {274} (\bibinfo {year}
  {2016})}\BibitemShut {NoStop}%
\bibitem [{\citenamefont {Wallerberger}\ \emph
  {et~al.}(2019{\natexlab{a}})\citenamefont {Wallerberger}, \citenamefont
  {Hausoel}, \citenamefont {Gunacker}, \citenamefont {Kowalski}, \citenamefont
  {Parragh}, \citenamefont {Goth}, \citenamefont {Held},\ and\ \citenamefont
  {Sangiovanni}}]{w2dynamics}%
  \BibitemOpen
  \bibfield  {author} {\bibinfo {author} {\bibfnamefont {M.}~\bibnamefont
  {Wallerberger}}, \bibinfo {author} {\bibfnamefont {A.}~\bibnamefont
  {Hausoel}}, \bibinfo {author} {\bibfnamefont {P.}~\bibnamefont {Gunacker}},
  \bibinfo {author} {\bibfnamefont {A.}~\bibnamefont {Kowalski}}, \bibinfo
  {author} {\bibfnamefont {N.}~\bibnamefont {Parragh}}, \bibinfo {author}
  {\bibfnamefont {F.}~\bibnamefont {Goth}}, \bibinfo {author} {\bibfnamefont
  {K.}~\bibnamefont {Held}},\ and\ \bibinfo {author} {\bibfnamefont
  {G.}~\bibnamefont {Sangiovanni}},\ }\bibfield  {title} {\bibinfo {title}
  {w2dynamics: Local one- and two-particle quantities from dynamical mean field
  theory},\ }\href {https://doi.org/https://doi.org/10.1016/j.cpc.2018.09.007}
  {\bibfield  {journal} {\bibinfo  {journal} {Computer Physics Communications}\
  }\textbf {\bibinfo {volume} {235}},\ \bibinfo {pages} {388 } (\bibinfo {year}
  {2019}{\natexlab{a}})}\BibitemShut {NoStop}%
\bibitem [{\citenamefont {Rubtsov}\ \emph {et~al.}(2008)\citenamefont
  {Rubtsov}, \citenamefont {Katsnelson},\ and\ \citenamefont
  {Lichtenstein}}]{Rubtsov08}%
  \BibitemOpen
  \bibfield  {author} {\bibinfo {author} {\bibfnamefont {A.~N.}\ \bibnamefont
  {Rubtsov}}, \bibinfo {author} {\bibfnamefont {M.~I.}\ \bibnamefont
  {Katsnelson}},\ and\ \bibinfo {author} {\bibfnamefont {A.~I.}\ \bibnamefont
  {Lichtenstein}},\ }\bibfield  {title} {\bibinfo {title} {Dual fermion
  approach to nonlocal correlations in the hubbard model},\ }\href
  {https://doi.org/10.1103/PhysRevB.77.033101} {\bibfield  {journal} {\bibinfo
  {journal} {Phys. Rev. B}\ }\textbf {\bibinfo {volume} {77}},\ \bibinfo
  {pages} {033101} (\bibinfo {year} {2008})}\BibitemShut {NoStop}%
\bibitem [{\citenamefont {Rohringer}\ \emph {et~al.}(2012)\citenamefont
  {Rohringer}, \citenamefont {Valli},\ and\ \citenamefont
  {Toschi}}]{Rohringer12}%
  \BibitemOpen
  \bibfield  {author} {\bibinfo {author} {\bibfnamefont {G.}~\bibnamefont
  {Rohringer}}, \bibinfo {author} {\bibfnamefont {A.}~\bibnamefont {Valli}},\
  and\ \bibinfo {author} {\bibfnamefont {A.}~\bibnamefont {Toschi}},\
  }\bibfield  {title} {\bibinfo {title} {Local electronic correlation at the
  two-particle level},\ }\href {https://doi.org/10.1103/PhysRevB.86.125114}
  {\bibfield  {journal} {\bibinfo  {journal} {Phys. Rev. B}\ }\textbf {\bibinfo
  {volume} {86}},\ \bibinfo {pages} {125114} (\bibinfo {year}
  {2012})}\BibitemShut {NoStop}%
\bibitem [{\citenamefont {Thunstr\"om}\ \emph {et~al.}(2018)\citenamefont
  {Thunstr\"om}, \citenamefont {Gunnarsson}, \citenamefont {Ciuchi},\ and\
  \citenamefont {Rohringer}}]{Thunstrom18}%
  \BibitemOpen
  \bibfield  {author} {\bibinfo {author} {\bibfnamefont {P.}~\bibnamefont
  {Thunstr\"om}}, \bibinfo {author} {\bibfnamefont {O.}~\bibnamefont
  {Gunnarsson}}, \bibinfo {author} {\bibfnamefont {S.}~\bibnamefont {Ciuchi}},\
  and\ \bibinfo {author} {\bibfnamefont {G.}~\bibnamefont {Rohringer}},\
  }\bibfield  {title} {\bibinfo {title} {Analytical investigation of
  singularities in two-particle irreducible vertex functions of the {Hubbard}
  atom},\ }\href {https://doi.org/10.1103/PhysRevB.98.235107} {\bibfield
  {journal} {\bibinfo  {journal} {Phys. Rev. B}\ }\textbf {\bibinfo {volume}
  {98}},\ \bibinfo {pages} {235107} (\bibinfo {year} {2018})}\BibitemShut
  {NoStop}%
\bibitem [{\citenamefont {Sidis}\ \emph {et~al.}(1999)\citenamefont {Sidis},
  \citenamefont {Braden}, \citenamefont {Bourges}, \citenamefont {Hennion},
  \citenamefont {NishiZaki}, \citenamefont {Maeno},\ and\ \citenamefont
  {Mori}}]{Sidis99}%
  \BibitemOpen
  \bibfield  {author} {\bibinfo {author} {\bibfnamefont {Y.}~\bibnamefont
  {Sidis}}, \bibinfo {author} {\bibfnamefont {M.}~\bibnamefont {Braden}},
  \bibinfo {author} {\bibfnamefont {P.}~\bibnamefont {Bourges}}, \bibinfo
  {author} {\bibfnamefont {B.}~\bibnamefont {Hennion}}, \bibinfo {author}
  {\bibfnamefont {S.}~\bibnamefont {NishiZaki}}, \bibinfo {author}
  {\bibfnamefont {Y.}~\bibnamefont {Maeno}},\ and\ \bibinfo {author}
  {\bibfnamefont {Y.}~\bibnamefont {Mori}},\ }\bibfield  {title} {\bibinfo
  {title} {Evidence for incommensurate spin fluctuations in
  {${\mathrm{Sr}}_{2}{\mathrm{RuO}}_{4}$}},\ }\href
  {https://doi.org/10.1103/PhysRevLett.83.3320} {\bibfield  {journal} {\bibinfo
   {journal} {Phys. Rev. Lett.}\ }\textbf {\bibinfo {volume} {83}},\ \bibinfo
  {pages} {3320} (\bibinfo {year} {1999})}\BibitemShut {NoStop}%
\bibitem [{\citenamefont {Steffens}\ \emph {et~al.}(2019)\citenamefont
  {Steffens}, \citenamefont {Sidis}, \citenamefont {Kulda}, \citenamefont
  {Mao}, \citenamefont {Maeno}, \citenamefont {Mazin},\ and\ \citenamefont
  {Braden}}]{Steffens19}%
  \BibitemOpen
  \bibfield  {author} {\bibinfo {author} {\bibfnamefont {P.}~\bibnamefont
  {Steffens}}, \bibinfo {author} {\bibfnamefont {Y.}~\bibnamefont {Sidis}},
  \bibinfo {author} {\bibfnamefont {J.}~\bibnamefont {Kulda}}, \bibinfo
  {author} {\bibfnamefont {Z.~Q.}\ \bibnamefont {Mao}}, \bibinfo {author}
  {\bibfnamefont {Y.}~\bibnamefont {Maeno}}, \bibinfo {author} {\bibfnamefont
  {I.~I.}\ \bibnamefont {Mazin}},\ and\ \bibinfo {author} {\bibfnamefont
  {M.}~\bibnamefont {Braden}},\ }\bibfield  {title} {\bibinfo {title} {Spin
  fluctuations in {${\mathrm{Sr}}_{2}{\mathrm{RuO}}_{4}$} from polarized
  neutron scattering: Implications for superconductivity},\ }\href
  {https://doi.org/10.1103/PhysRevLett.122.047004} {\bibfield  {journal}
  {\bibinfo  {journal} {Phys. Rev. Lett.}\ }\textbf {\bibinfo {volume} {122}},\
  \bibinfo {pages} {047004} (\bibinfo {year} {2019})}\BibitemShut {NoStop}%
\bibitem [{\citenamefont {Boehnke}\ \emph {et~al.}(2018)\citenamefont
  {Boehnke}, \citenamefont {Werner},\ and\ \citenamefont
  {Lechermann}}]{0295-5075-122-5-57001}%
  \BibitemOpen
  \bibfield  {author} {\bibinfo {author} {\bibfnamefont {L.}~\bibnamefont
  {Boehnke}}, \bibinfo {author} {\bibfnamefont {P.}~\bibnamefont {Werner}},\
  and\ \bibinfo {author} {\bibfnamefont {F.}~\bibnamefont {Lechermann}},\
  }\bibfield  {title} {\bibinfo {title} {Multi-orbital nature of the spin
  fluctuations in {Sr$_2$RuO$_4$}},\ }\href@noop {} {\bibfield  {journal}
  {\bibinfo  {journal} {EPL}\ }\textbf {\bibinfo {volume} {122}},\ \bibinfo
  {pages} {57001} (\bibinfo {year} {2018})}\BibitemShut {NoStop}%
\bibitem [{\citenamefont {Acharya}\ \emph {et~al.}(2019)\citenamefont
  {Acharya}, \citenamefont {Pashov}, \citenamefont {Weber}, \citenamefont
  {Park}, \citenamefont {Sponza},\ and\ \citenamefont
  {Schilfgaarde}}]{Acharya:2019aa}%
  \BibitemOpen
  \bibfield  {author} {\bibinfo {author} {\bibfnamefont {S.}~\bibnamefont
  {Acharya}}, \bibinfo {author} {\bibfnamefont {D.}~\bibnamefont {Pashov}},
  \bibinfo {author} {\bibfnamefont {C.}~\bibnamefont {Weber}}, \bibinfo
  {author} {\bibfnamefont {H.}~\bibnamefont {Park}}, \bibinfo {author}
  {\bibfnamefont {L.}~\bibnamefont {Sponza}},\ and\ \bibinfo {author}
  {\bibfnamefont {M.~V.}\ \bibnamefont {Schilfgaarde}},\ }\bibfield  {title}
  {\bibinfo {title} {Evening out the spin and charge parity to increase {$T_c$}
  in {Sr$_2$RuO$_4$}},\ }\href@noop {} {\bibfield  {journal} {\bibinfo
  {journal} {Communications Physics}\ }\textbf {\bibinfo {volume} {2}},\
  \bibinfo {pages} {163} (\bibinfo {year} {2019})}\BibitemShut {NoStop}%
\bibitem [{\citenamefont {Mostofi}\ \emph {et~al.}(2008)\citenamefont
  {Mostofi}, \citenamefont {Yates}, \citenamefont {Lee}, \citenamefont {Souza},
  \citenamefont {Vanderbilt},\ and\ \citenamefont {Marzari}}]{Mostofi:2008aa}%
  \BibitemOpen
  \bibfield  {author} {\bibinfo {author} {\bibfnamefont {A.~A.}\ \bibnamefont
  {Mostofi}}, \bibinfo {author} {\bibfnamefont {J.~R.}\ \bibnamefont {Yates}},
  \bibinfo {author} {\bibfnamefont {Y.-S.}\ \bibnamefont {Lee}}, \bibinfo
  {author} {\bibfnamefont {I.}~\bibnamefont {Souza}}, \bibinfo {author}
  {\bibfnamefont {D.}~\bibnamefont {Vanderbilt}},\ and\ \bibinfo {author}
  {\bibfnamefont {N.}~\bibnamefont {Marzari}},\ }\bibfield  {title} {\bibinfo
  {title} {wannier90: A tool for obtaining maximally-localised {Wannier}
  functions},\ }\href@noop {} {\bibfield  {journal} {\bibinfo  {journal}
  {Comput. Phys. Commun.}\ }\textbf {\bibinfo {volume} {178}},\ \bibinfo
  {pages} {685} (\bibinfo {year} {2008})}\BibitemShut {NoStop}%
\bibitem [{\citenamefont {Mortensen}\ \emph {et~al.}(2005)\citenamefont
  {Mortensen}, \citenamefont {Hansen},\ and\ \citenamefont
  {Jacobsen}}]{Mortensen2005}%
  \BibitemOpen
  \bibfield  {author} {\bibinfo {author} {\bibfnamefont {J.~J.}\ \bibnamefont
  {Mortensen}}, \bibinfo {author} {\bibfnamefont {L.~B.}\ \bibnamefont
  {Hansen}},\ and\ \bibinfo {author} {\bibfnamefont {K.~W.}\ \bibnamefont
  {Jacobsen}},\ }\bibfield  {title} {\bibinfo {title} {Real-space grid
  implementation of the projector augmented wave method},\ }\href
  {https://doi.org/10.1103/PhysRevB.71.035109} {\bibfield  {journal} {\bibinfo
  {journal} {Phys. Rev. B}\ }\textbf {\bibinfo {volume} {71}},\ \bibinfo
  {pages} {035109} (\bibinfo {year} {2005})}\BibitemShut {NoStop}%
\bibitem [{\citenamefont {Enkovaara}\ \emph {et~al.}(2010)\citenamefont
  {Enkovaara}, \citenamefont {Rostgaard}, \citenamefont {Mortensen},
  \citenamefont {Chen}, \citenamefont {Du{\l}ak}, \citenamefont {Ferrighi},
  \citenamefont {Gavnholt}, \citenamefont {Glinsvad}, \citenamefont {Haikola},
  \citenamefont {Hansen}, \citenamefont {Kristoffersen}, \citenamefont
  {Kuisma}, \citenamefont {Larsen}, \citenamefont {Lehtovaara}, \citenamefont
  {Ljungberg}, \citenamefont {Lopez-Acevedo}, \citenamefont {Moses},
  \citenamefont {Ojanen}, \citenamefont {Olsen}, \citenamefont {Petzold},
  \citenamefont {Romero}, \citenamefont {Stausholm-M{\o}ller}, \citenamefont
  {Strange}, \citenamefont {Tritsaris}, \citenamefont {Vanin}, \citenamefont
  {Walter}, \citenamefont {Hammer}, \citenamefont {H{\"a}kkinen}, \citenamefont
  {Madsen}, \citenamefont {Nieminen}, \citenamefont {N{\o}rskov}, \citenamefont
  {Puska}, \citenamefont {Rantala}, \citenamefont {Schi{\o}tz}, \citenamefont
  {Thygesen},\ and\ \citenamefont {Jacobsen}}]{Enkovaara_2010}%
  \BibitemOpen
  \bibfield  {author} {\bibinfo {author} {\bibfnamefont {J.}~\bibnamefont
  {Enkovaara}}, \bibinfo {author} {\bibfnamefont {C.}~\bibnamefont
  {Rostgaard}}, \bibinfo {author} {\bibfnamefont {J.~J.}\ \bibnamefont
  {Mortensen}}, \bibinfo {author} {\bibfnamefont {J.}~\bibnamefont {Chen}},
  \bibinfo {author} {\bibfnamefont {M.}~\bibnamefont {Du{\l}ak}}, \bibinfo
  {author} {\bibfnamefont {L.}~\bibnamefont {Ferrighi}}, \bibinfo {author}
  {\bibfnamefont {J.}~\bibnamefont {Gavnholt}}, \bibinfo {author}
  {\bibfnamefont {C.}~\bibnamefont {Glinsvad}}, \bibinfo {author}
  {\bibfnamefont {V.}~\bibnamefont {Haikola}}, \bibinfo {author} {\bibfnamefont
  {H.~A.}\ \bibnamefont {Hansen}}, \bibinfo {author} {\bibfnamefont {H.~H.}\
  \bibnamefont {Kristoffersen}}, \bibinfo {author} {\bibfnamefont
  {M.}~\bibnamefont {Kuisma}}, \bibinfo {author} {\bibfnamefont {A.~H.}\
  \bibnamefont {Larsen}}, \bibinfo {author} {\bibfnamefont {L.}~\bibnamefont
  {Lehtovaara}}, \bibinfo {author} {\bibfnamefont {M.}~\bibnamefont
  {Ljungberg}}, \bibinfo {author} {\bibfnamefont {O.}~\bibnamefont
  {Lopez-Acevedo}}, \bibinfo {author} {\bibfnamefont {P.~G.}\ \bibnamefont
  {Moses}}, \bibinfo {author} {\bibfnamefont {J.}~\bibnamefont {Ojanen}},
  \bibinfo {author} {\bibfnamefont {T.}~\bibnamefont {Olsen}}, \bibinfo
  {author} {\bibfnamefont {V.}~\bibnamefont {Petzold}}, \bibinfo {author}
  {\bibfnamefont {N.~A.}\ \bibnamefont {Romero}}, \bibinfo {author}
  {\bibfnamefont {J.}~\bibnamefont {Stausholm-M{\o}ller}}, \bibinfo {author}
  {\bibfnamefont {M.}~\bibnamefont {Strange}}, \bibinfo {author} {\bibfnamefont
  {G.~A.}\ \bibnamefont {Tritsaris}}, \bibinfo {author} {\bibfnamefont
  {M.}~\bibnamefont {Vanin}}, \bibinfo {author} {\bibfnamefont
  {M.}~\bibnamefont {Walter}}, \bibinfo {author} {\bibfnamefont
  {B.}~\bibnamefont {Hammer}}, \bibinfo {author} {\bibfnamefont
  {H.}~\bibnamefont {H{\"a}kkinen}}, \bibinfo {author} {\bibfnamefont
  {G.~K.~H.}\ \bibnamefont {Madsen}}, \bibinfo {author} {\bibfnamefont {R.~M.}\
  \bibnamefont {Nieminen}}, \bibinfo {author} {\bibfnamefont {J.~K.}\
  \bibnamefont {N{\o}rskov}}, \bibinfo {author} {\bibfnamefont
  {M.}~\bibnamefont {Puska}}, \bibinfo {author} {\bibfnamefont {T.~T.}\
  \bibnamefont {Rantala}}, \bibinfo {author} {\bibfnamefont {J.}~\bibnamefont
  {Schi{\o}tz}}, \bibinfo {author} {\bibfnamefont {K.~S.}\ \bibnamefont
  {Thygesen}},\ and\ \bibinfo {author} {\bibfnamefont {K.~W.}\ \bibnamefont
  {Jacobsen}},\ }\bibfield  {title} {\bibinfo {title} {Electronic structure
  calculations with {GPAW}: a real-space implementation of the projector
  augmented-wave method},\ }\href@noop {} {\bibfield  {journal} {\bibinfo
  {journal} {J. Phys.: Condens. Matter}\ }\textbf {\bibinfo {volume} {22}},\
  \bibinfo {pages} {253202} (\bibinfo {year} {2010})}\BibitemShut {NoStop}%
\bibitem [{\citenamefont {Perdew}\ \emph {et~al.}(1996)\citenamefont {Perdew},
  \citenamefont {Burke},\ and\ \citenamefont
  {Ernzerhof}}]{PhysRevLett.77.3865}%
  \BibitemOpen
  \bibfield  {author} {\bibinfo {author} {\bibfnamefont {J.~P.}\ \bibnamefont
  {Perdew}}, \bibinfo {author} {\bibfnamefont {K.}~\bibnamefont {Burke}},\ and\
  \bibinfo {author} {\bibfnamefont {M.}~\bibnamefont {Ernzerhof}},\ }\bibfield
  {title} {\bibinfo {title} {Generalized gradient approximation made simple},\
  }\href@noop {} {\bibfield  {journal} {\bibinfo  {journal} {Phys. Rev. Lett.}\
  }\textbf {\bibinfo {volume} {77}},\ \bibinfo {pages} {3865} (\bibinfo {year}
  {1996})}\BibitemShut {NoStop}%
\bibitem [{\citenamefont {Methfessel}\ and\ \citenamefont
  {Paxton}(1989)}]{PhysRevB.40.3616}%
  \BibitemOpen
  \bibfield  {author} {\bibinfo {author} {\bibfnamefont {M.}~\bibnamefont
  {Methfessel}}\ and\ \bibinfo {author} {\bibfnamefont {A.~T.}\ \bibnamefont
  {Paxton}},\ }\bibfield  {title} {\bibinfo {title} {High-precision sampling
  for {Brillouin-zone} integration in metals},\ }\href@noop {} {\bibfield
  {journal} {\bibinfo  {journal} {Phys. Rev. B}\ }\textbf {\bibinfo {volume}
  {40}},\ \bibinfo {pages} {3616} (\bibinfo {year} {1989})}\BibitemShut
  {NoStop}%
\bibitem [{\citenamefont {Vogt}\ and\ \citenamefont
  {Buttrey}(1995)}]{PhysRevB.52.R9843}%
  \BibitemOpen
  \bibfield  {author} {\bibinfo {author} {\bibfnamefont {T.}~\bibnamefont
  {Vogt}}\ and\ \bibinfo {author} {\bibfnamefont {D.~J.}\ \bibnamefont
  {Buttrey}},\ }\bibfield  {title} {\bibinfo {title} {Low-temperature
  structural behavior of {${\mathrm{Sr}}_{2}$${\mathrm{RuO}}_{4}$}},\
  }\href@noop {} {\bibfield  {journal} {\bibinfo  {journal} {Phys. Rev. B}\
  }\textbf {\bibinfo {volume} {52}},\ \bibinfo {pages} {R9843} (\bibinfo {year}
  {1995})}\BibitemShut {NoStop}%
\bibitem [{\citenamefont {Wallerberger}\ \emph
  {et~al.}(2019{\natexlab{b}})\citenamefont {Wallerberger}, \citenamefont
  {Hausoel}, \citenamefont {Gunacker}, \citenamefont {Kowalski}, \citenamefont
  {Parragh}, \citenamefont {Goth}, \citenamefont {Held},\ and\ \citenamefont
  {Sangiovanni}}]{WALLERBERGER2019388}%
  \BibitemOpen
  \bibfield  {author} {\bibinfo {author} {\bibfnamefont {M.}~\bibnamefont
  {Wallerberger}}, \bibinfo {author} {\bibfnamefont {A.}~\bibnamefont
  {Hausoel}}, \bibinfo {author} {\bibfnamefont {P.}~\bibnamefont {Gunacker}},
  \bibinfo {author} {\bibfnamefont {A.}~\bibnamefont {Kowalski}}, \bibinfo
  {author} {\bibfnamefont {N.}~\bibnamefont {Parragh}}, \bibinfo {author}
  {\bibfnamefont {F.}~\bibnamefont {Goth}}, \bibinfo {author} {\bibfnamefont
  {K.}~\bibnamefont {Held}},\ and\ \bibinfo {author} {\bibfnamefont
  {G.}~\bibnamefont {Sangiovanni}},\ }\bibfield  {title} {\bibinfo {title}
  {w2dynamics: Local one- and two-particle quantities from dynamical mean field
  theory},\ }\href@noop {} {\bibfield  {journal} {\bibinfo  {journal} {Comput.
  Phys. Commun.}\ }\textbf {\bibinfo {volume} {235}},\ \bibinfo {pages} {388}
  (\bibinfo {year} {2019}{\natexlab{b}})}\BibitemShut {NoStop}%
\bibitem [{\citenamefont {Gunacker}\ \emph {et~al.}(2015)\citenamefont
  {Gunacker}, \citenamefont {Wallerberger}, \citenamefont {Gull}, \citenamefont
  {Hausoel}, \citenamefont {Sangiovanni},\ and\ \citenamefont
  {Held}}]{PhysRevB.92.155102}%
  \BibitemOpen
  \bibfield  {author} {\bibinfo {author} {\bibfnamefont {P.}~\bibnamefont
  {Gunacker}}, \bibinfo {author} {\bibfnamefont {M.}~\bibnamefont
  {Wallerberger}}, \bibinfo {author} {\bibfnamefont {E.}~\bibnamefont {Gull}},
  \bibinfo {author} {\bibfnamefont {A.}~\bibnamefont {Hausoel}}, \bibinfo
  {author} {\bibfnamefont {G.}~\bibnamefont {Sangiovanni}},\ and\ \bibinfo
  {author} {\bibfnamefont {K.}~\bibnamefont {Held}},\ }\bibfield  {title}
  {\bibinfo {title} {Continuous-time quantum {Monte Carlo} using worm
  sampling},\ }\href@noop {} {\bibfield  {journal} {\bibinfo  {journal} {Phys.
  Rev. B}\ }\textbf {\bibinfo {volume} {92}},\ \bibinfo {pages} {155102}
  (\bibinfo {year} {2015})}\BibitemShut {NoStop}%
\bibitem [{\citenamefont {Nekrasov}\ \emph {et~al.}(2006)\citenamefont
  {Nekrasov}, \citenamefont {Held}, \citenamefont {Keller}, \citenamefont
  {Kondakov}, \citenamefont {Pruschke}, \citenamefont {Kollar}, \citenamefont
  {Andersen}, \citenamefont {Anisimov},\ and\ \citenamefont
  {Vollhardt}}]{Nekrasov06}%
  \BibitemOpen
  \bibfield  {author} {\bibinfo {author} {\bibfnamefont {I.~A.}\ \bibnamefont
  {Nekrasov}}, \bibinfo {author} {\bibfnamefont {K.}~\bibnamefont {Held}},
  \bibinfo {author} {\bibfnamefont {G.}~\bibnamefont {Keller}}, \bibinfo
  {author} {\bibfnamefont {D.~E.}\ \bibnamefont {Kondakov}}, \bibinfo {author}
  {\bibfnamefont {T.}~\bibnamefont {Pruschke}}, \bibinfo {author}
  {\bibfnamefont {M.}~\bibnamefont {Kollar}}, \bibinfo {author} {\bibfnamefont
  {O.~K.}\ \bibnamefont {Andersen}}, \bibinfo {author} {\bibfnamefont {V.~I.}\
  \bibnamefont {Anisimov}},\ and\ \bibinfo {author} {\bibfnamefont
  {D.}~\bibnamefont {Vollhardt}},\ }\bibfield  {title} {\bibinfo {title}
  {Momentum-resolved spectral functions of {${\mathrm{SrVO}}_{3}$} calculated
  by $\mathrm{LDA}+\mathrm{DMFT}$},\ }\href
  {https://doi.org/10.1103/PhysRevB.73.155112} {\bibfield  {journal} {\bibinfo
  {journal} {Phys. Rev. B}\ }\textbf {\bibinfo {volume} {73}},\ \bibinfo
  {pages} {155112} (\bibinfo {year} {2006})}\BibitemShut {NoStop}%
\bibitem [{\citenamefont {Sakuma}\ \emph {et~al.}(2013)\citenamefont {Sakuma},
  \citenamefont {Werner},\ and\ \citenamefont {Aryasetiawan}}]{Sakuma13}%
  \BibitemOpen
  \bibfield  {author} {\bibinfo {author} {\bibfnamefont {R.}~\bibnamefont
  {Sakuma}}, \bibinfo {author} {\bibfnamefont {P.}~\bibnamefont {Werner}},\
  and\ \bibinfo {author} {\bibfnamefont {F.}~\bibnamefont {Aryasetiawan}},\
  }\bibfield  {title} {\bibinfo {title} {Electronic structure of {SrVO}${}_{3}$
  within {$GW$}+{DMFT}},\ }\href {https://doi.org/10.1103/PhysRevB.88.235110}
  {\bibfield  {journal} {\bibinfo  {journal} {Phys. Rev. B}\ }\textbf {\bibinfo
  {volume} {88}},\ \bibinfo {pages} {235110} (\bibinfo {year}
  {2013})}\BibitemShut {NoStop}%
\bibitem [{\citenamefont {Zhang}\ \emph {et~al.}(2016)\citenamefont {Zhang},
  \citenamefont {Zhou}, \citenamefont {Guo}, \citenamefont {Zhao},
  \citenamefont {Barnes}, \citenamefont {Zhang}, \citenamefont {Eaton},
  \citenamefont {Zheng}, \citenamefont {Brahlek}, \citenamefont {Haneef} \emph
  {et~al.}}]{zhang2016correlated}%
  \BibitemOpen
  \bibfield  {author} {\bibinfo {author} {\bibfnamefont {L.}~\bibnamefont
  {Zhang}}, \bibinfo {author} {\bibfnamefont {Y.}~\bibnamefont {Zhou}},
  \bibinfo {author} {\bibfnamefont {L.}~\bibnamefont {Guo}}, \bibinfo {author}
  {\bibfnamefont {W.}~\bibnamefont {Zhao}}, \bibinfo {author} {\bibfnamefont
  {A.}~\bibnamefont {Barnes}}, \bibinfo {author} {\bibfnamefont {H.-T.}\
  \bibnamefont {Zhang}}, \bibinfo {author} {\bibfnamefont {C.}~\bibnamefont
  {Eaton}}, \bibinfo {author} {\bibfnamefont {Y.}~\bibnamefont {Zheng}},
  \bibinfo {author} {\bibfnamefont {M.}~\bibnamefont {Brahlek}}, \bibinfo
  {author} {\bibfnamefont {H.~F.}\ \bibnamefont {Haneef}}, \emph {et~al.},\
  }\bibfield  {title} {\bibinfo {title} {Correlated metals as transparent
  conductors},\ }\href@noop {} {\bibfield  {journal} {\bibinfo  {journal}
  {Nature materials}\ }\textbf {\bibinfo {volume} {15}},\ \bibinfo {pages}
  {204} (\bibinfo {year} {2016})}\BibitemShut {NoStop}%
\bibitem [{\citenamefont {Galler}\ \emph {et~al.}(2017)\citenamefont {Galler},
  \citenamefont {Thunstr\"om}, \citenamefont {Gunacker}, \citenamefont
  {Tomczak},\ and\ \citenamefont {Held}}]{Galler17}%
  \BibitemOpen
  \bibfield  {author} {\bibinfo {author} {\bibfnamefont {A.}~\bibnamefont
  {Galler}}, \bibinfo {author} {\bibfnamefont {P.}~\bibnamefont {Thunstr\"om}},
  \bibinfo {author} {\bibfnamefont {P.}~\bibnamefont {Gunacker}}, \bibinfo
  {author} {\bibfnamefont {J.~M.}\ \bibnamefont {Tomczak}},\ and\ \bibinfo
  {author} {\bibfnamefont {K.}~\bibnamefont {Held}},\ }\bibfield  {title}
  {\bibinfo {title} {Ab initio dynamical vertex approximation},\ }\href
  {https://doi.org/10.1103/PhysRevB.95.115107} {\bibfield  {journal} {\bibinfo
  {journal} {Phys. Rev. B}\ }\textbf {\bibinfo {volume} {95}},\ \bibinfo
  {pages} {115107} (\bibinfo {year} {2017})}\BibitemShut {NoStop}%
\bibitem [{\citenamefont {Khurana}(1990)}]{Khurana90}%
  \BibitemOpen
  \bibfield  {author} {\bibinfo {author} {\bibfnamefont {A.}~\bibnamefont
  {Khurana}},\ }\bibfield  {title} {\bibinfo {title} {Electrical conductivity
  in the infinite-dimensional {H}ubbard model},\ }\href
  {https://doi.org/10.1103/PhysRevLett.64.1990} {\bibfield  {journal} {\bibinfo
   {journal} {Phys. Rev. Lett.}\ }\textbf {\bibinfo {volume} {64}},\ \bibinfo
  {pages} {1990} (\bibinfo {year} {1990})}\BibitemShut {NoStop}%
\bibitem [{\citenamefont {Schweitzer}\ and\ \citenamefont
  {Czycholl}(1991)}]{Schweitzer91}%
  \BibitemOpen
  \bibfield  {author} {\bibinfo {author} {\bibfnamefont {H.}~\bibnamefont
  {Schweitzer}}\ and\ \bibinfo {author} {\bibfnamefont {G.}~\bibnamefont
  {Czycholl}},\ }\bibfield  {title} {\bibinfo {title} {Resistivity and
  thermopower of heavy-fermion systems},\ }\href
  {https://doi.org/10.1103/PhysRevLett.67.3724} {\bibfield  {journal} {\bibinfo
   {journal} {Phys. Rev. Lett.}\ }\textbf {\bibinfo {volume} {67}},\ \bibinfo
  {pages} {3724} (\bibinfo {year} {1991})}\BibitemShut {NoStop}%
\bibitem [{\citenamefont {Jarrell}(1992)}]{PhysRevLett.69.168}%
  \BibitemOpen
  \bibfield  {author} {\bibinfo {author} {\bibfnamefont {M.}~\bibnamefont
  {Jarrell}},\ }\bibfield  {title} {\bibinfo {title} {Hubbard model in infinite
  dimensions: A quantum {Monte Carlo} study},\ }\href@noop {} {\bibfield
  {journal} {\bibinfo  {journal} {Phys. Rev. Lett.}\ }\textbf {\bibinfo
  {volume} {69}},\ \bibinfo {pages} {168} (\bibinfo {year} {1992})}\BibitemShut
  {NoStop}%
\bibitem [{\citenamefont {Brener}\ \emph {et~al.}(2008)\citenamefont {Brener},
  \citenamefont {Hafermann}, \citenamefont {Rubtsov}, \citenamefont
  {Katsnelson},\ and\ \citenamefont {Lichtenstein}}]{Brener08}%
  \BibitemOpen
  \bibfield  {author} {\bibinfo {author} {\bibfnamefont {S.}~\bibnamefont
  {Brener}}, \bibinfo {author} {\bibfnamefont {H.}~\bibnamefont {Hafermann}},
  \bibinfo {author} {\bibfnamefont {A.~N.}\ \bibnamefont {Rubtsov}}, \bibinfo
  {author} {\bibfnamefont {M.~I.}\ \bibnamefont {Katsnelson}},\ and\ \bibinfo
  {author} {\bibfnamefont {A.~I.}\ \bibnamefont {Lichtenstein}},\ }\bibfield
  {title} {\bibinfo {title} {Dual fermion approach to susceptibility of
  correlated lattice fermions},\ }\href
  {https://doi.org/10.1103/PhysRevB.77.195105} {\bibfield  {journal} {\bibinfo
  {journal} {Phys. Rev. B}\ }\textbf {\bibinfo {volume} {77}},\ \bibinfo
  {pages} {195105} (\bibinfo {year} {2008})}\BibitemShut {NoStop}%
\bibitem [{\citenamefont {Brener}\ \emph {et~al.}(2020)\citenamefont {Brener},
  \citenamefont {Stepanov}, \citenamefont {Rubtsov}, \citenamefont
  {Katsnelson},\ and\ \citenamefont {Lichtenstein}}]{Brener20}%
  \BibitemOpen
  \bibfield  {author} {\bibinfo {author} {\bibfnamefont {S.}~\bibnamefont
  {Brener}}, \bibinfo {author} {\bibfnamefont {E.~A.}\ \bibnamefont
  {Stepanov}}, \bibinfo {author} {\bibfnamefont {A.~N.}\ \bibnamefont
  {Rubtsov}}, \bibinfo {author} {\bibfnamefont {M.~I.}\ \bibnamefont
  {Katsnelson}},\ and\ \bibinfo {author} {\bibfnamefont {A.~I.}\ \bibnamefont
  {Lichtenstein}},\ }\bibfield  {title} {\bibinfo {title} {Dual fermion method
  as a prototype of generic reference-system approach for correlated
  fermions},\ }\href
  {https://doi.org/https://doi.org/10.1016/j.aop.2020.168310} {\bibfield
  {journal} {\bibinfo  {journal} {Annals of Physics}\ }\textbf {\bibinfo
  {volume} {422}},\ \bibinfo {pages} {168310} (\bibinfo {year}
  {2020})}\BibitemShut {NoStop}%
\bibitem [{\citenamefont {Galler}\ \emph {et~al.}(2018)\citenamefont {Galler},
  \citenamefont {Kaufmann}, \citenamefont {Gunacker}, \citenamefont {Pickem},
  \citenamefont {Thunstr\"{o}m}, \citenamefont {Tomczak},\ and\ \citenamefont
  {Held}}]{Galler18}%
  \BibitemOpen
  \bibfield  {author} {\bibinfo {author} {\bibfnamefont {A.}~\bibnamefont
  {Galler}}, \bibinfo {author} {\bibfnamefont {J.}~\bibnamefont {Kaufmann}},
  \bibinfo {author} {\bibfnamefont {P.}~\bibnamefont {Gunacker}}, \bibinfo
  {author} {\bibfnamefont {M.}~\bibnamefont {Pickem}}, \bibinfo {author}
  {\bibfnamefont {P.}~\bibnamefont {Thunstr\"{o}m}}, \bibinfo {author}
  {\bibfnamefont {J.~M.}\ \bibnamefont {Tomczak}},\ and\ \bibinfo {author}
  {\bibfnamefont {K.}~\bibnamefont {Held}},\ }\bibfield  {title} {\bibinfo
  {title} {Towards ab initio calculations with the dynamical vertex
  approximation},\ }\href {https://doi.org/10.7566/JPSJ.87.041004} {\bibfield
  {journal} {\bibinfo  {journal} {Journal of the Physical Society of Japan}\
  }\textbf {\bibinfo {volume} {87}},\ \bibinfo {pages} {041004} (\bibinfo
  {year} {2018})},\ \Eprint
  {https://arxiv.org/abs/https://doi.org/10.7566/JPSJ.87.041004}
  {https://doi.org/10.7566/JPSJ.87.041004} \BibitemShut {NoStop}%
\bibitem [{\citenamefont {Shinaoka}\ \emph {et~al.}(2018)\citenamefont
  {Shinaoka}, \citenamefont {Otsuki}, \citenamefont {Haule}, \citenamefont
  {Wallerberger}, \citenamefont {Gull}, \citenamefont {Yoshimi},\ and\
  \citenamefont {Ohzeki}}]{PhysRevB.97.205111}%
  \BibitemOpen
  \bibfield  {author} {\bibinfo {author} {\bibfnamefont {H.}~\bibnamefont
  {Shinaoka}}, \bibinfo {author} {\bibfnamefont {J.}~\bibnamefont {Otsuki}},
  \bibinfo {author} {\bibfnamefont {K.}~\bibnamefont {Haule}}, \bibinfo
  {author} {\bibfnamefont {M.}~\bibnamefont {Wallerberger}}, \bibinfo {author}
  {\bibfnamefont {E.}~\bibnamefont {Gull}}, \bibinfo {author} {\bibfnamefont
  {K.}~\bibnamefont {Yoshimi}},\ and\ \bibinfo {author} {\bibfnamefont
  {M.}~\bibnamefont {Ohzeki}},\ }\bibfield  {title} {\bibinfo {title}
  {Overcomplete compact representation of two-particle green's functions},\
  }\href@noop {} {\bibfield  {journal} {\bibinfo  {journal} {Phys. Rev. B}\
  }\textbf {\bibinfo {volume} {97}},\ \bibinfo {pages} {205111} (\bibinfo
  {year} {2018})}\BibitemShut {NoStop}%
\bibitem [{\citenamefont {Wallerberger}\ \emph {et~al.}(2021)\citenamefont
  {Wallerberger}, \citenamefont {Shinaoka},\ and\ \citenamefont
  {Kauch}}]{PhysRevResearch.3.033168}%
  \BibitemOpen
  \bibfield  {author} {\bibinfo {author} {\bibfnamefont {M.}~\bibnamefont
  {Wallerberger}}, \bibinfo {author} {\bibfnamefont {H.}~\bibnamefont
  {Shinaoka}},\ and\ \bibinfo {author} {\bibfnamefont {A.}~\bibnamefont
  {Kauch}},\ }\bibfield  {title} {\bibinfo {title} {Solving the
  {Bethe-Salpeter} equation with exponential convergence},\ }\href@noop {}
  {\bibfield  {journal} {\bibinfo  {journal} {Phys. Rev. Res.}\ }\textbf
  {\bibinfo {volume} {3}},\ \bibinfo {pages} {033168} (\bibinfo {year}
  {2021})}\BibitemShut {NoStop}%
\bibitem [{\citenamefont {van Loon}\ and\ \citenamefont
  {Strand}(2023)}]{vanloon2023larmor}%
  \BibitemOpen
  \bibfield  {author} {\bibinfo {author} {\bibfnamefont {E.~G. C.~P.}\
  \bibnamefont {van Loon}}\ and\ \bibinfo {author} {\bibfnamefont {H.~U.~R.}\
  \bibnamefont {Strand}},\ }\bibfield  {title} {\bibinfo {title} {Larmor
  precession in strongly correlated itinerant electron systems},\ }\href
  {https://doi.org/10.1038/s42005-023-01411-w} {\bibfield  {journal} {\bibinfo
  {journal} {Communications Physics}\ }\textbf {\bibinfo {volume} {6}},\
  \bibinfo {pages} {289} (\bibinfo {year} {2023})}\BibitemShut {NoStop}%
\bibitem [{\citenamefont {Hausoel}\ \emph {et~al.}(2017)\citenamefont
  {Hausoel}, \citenamefont {Karolak}, \citenamefont
  {{\c{S}}a{\c{s}}$\iota$o{\u{g}}lu}, \citenamefont {Lichtenstein},
  \citenamefont {Held}, \citenamefont {Katanin}, \citenamefont {Toschi},\ and\
  \citenamefont {Sangiovanni}}]{hausoel2017local}%
  \BibitemOpen
  \bibfield  {author} {\bibinfo {author} {\bibfnamefont {A.}~\bibnamefont
  {Hausoel}}, \bibinfo {author} {\bibfnamefont {M.}~\bibnamefont {Karolak}},
  \bibinfo {author} {\bibfnamefont {E.}~\bibnamefont
  {{\c{S}}a{\c{s}}$\iota$o{\u{g}}lu}}, \bibinfo {author} {\bibfnamefont
  {A.}~\bibnamefont {Lichtenstein}}, \bibinfo {author} {\bibfnamefont
  {K.}~\bibnamefont {Held}}, \bibinfo {author} {\bibfnamefont {A.}~\bibnamefont
  {Katanin}}, \bibinfo {author} {\bibfnamefont {A.}~\bibnamefont {Toschi}},\
  and\ \bibinfo {author} {\bibfnamefont {G.}~\bibnamefont {Sangiovanni}},\
  }\bibfield  {title} {\bibinfo {title} {Local magnetic moments in iron and
  nickel at ambient and earth’s core conditions},\ }\href@noop {} {\bibfield
  {journal} {\bibinfo  {journal} {Nature communications}\ }\textbf {\bibinfo
  {volume} {8}},\ \bibinfo {pages} {16062} (\bibinfo {year}
  {2017})}\BibitemShut {NoStop}%
\bibitem [{\citenamefont {Geffroy}\ \emph {et~al.}(2019)\citenamefont
  {Geffroy}, \citenamefont {Kaufmann}, \citenamefont {Hariki}, \citenamefont
  {Gunacker}, \citenamefont {Hausoel},\ and\ \citenamefont
  {Kune\ifmmode~\check{s}\else \v{s}\fi{}}}]{Geffroy19}%
  \BibitemOpen
  \bibfield  {author} {\bibinfo {author} {\bibfnamefont {D.}~\bibnamefont
  {Geffroy}}, \bibinfo {author} {\bibfnamefont {J.}~\bibnamefont {Kaufmann}},
  \bibinfo {author} {\bibfnamefont {A.}~\bibnamefont {Hariki}}, \bibinfo
  {author} {\bibfnamefont {P.}~\bibnamefont {Gunacker}}, \bibinfo {author}
  {\bibfnamefont {A.}~\bibnamefont {Hausoel}},\ and\ \bibinfo {author}
  {\bibfnamefont {J.}~\bibnamefont {Kune\ifmmode~\check{s}\else \v{s}\fi{}}},\
  }\bibfield  {title} {\bibinfo {title} {Collective modes in excitonic magnets:
  Dynamical mean-field study},\ }\href
  {https://doi.org/10.1103/PhysRevLett.122.127601} {\bibfield  {journal}
  {\bibinfo  {journal} {Phys. Rev. Lett.}\ }\textbf {\bibinfo {volume} {122}},\
  \bibinfo {pages} {127601} (\bibinfo {year} {2019})}\BibitemShut {NoStop}%
\bibitem [{\citenamefont {Pickem}\ \emph {et~al.}(2021)\citenamefont {Pickem},
  \citenamefont {Kaufmann}, \citenamefont {Held},\ and\ \citenamefont
  {Tomczak}}]{Pickem21}%
  \BibitemOpen
  \bibfield  {author} {\bibinfo {author} {\bibfnamefont {M.}~\bibnamefont
  {Pickem}}, \bibinfo {author} {\bibfnamefont {J.}~\bibnamefont {Kaufmann}},
  \bibinfo {author} {\bibfnamefont {K.}~\bibnamefont {Held}},\ and\ \bibinfo
  {author} {\bibfnamefont {J.~M.}\ \bibnamefont {Tomczak}},\ }\bibfield
  {title} {\bibinfo {title} {Zoology of spin and orbital fluctuations in
  ultrathin oxide films},\ }\href {https://doi.org/10.1103/PhysRevB.104.024307}
  {\bibfield  {journal} {\bibinfo  {journal} {Phys. Rev. B}\ }\textbf {\bibinfo
  {volume} {104}},\ \bibinfo {pages} {024307} (\bibinfo {year}
  {2021})}\BibitemShut {NoStop}%
\end{thebibliography}%

\end{document}